\documentclass[usenatbib]{mn2e}  
\usepackage{graphicx}
\usepackage{txfonts}
\usepackage{multirow}
\usepackage{supertabular}
\usepackage{longtable}
\usepackage{rotating}
\usepackage{amssymb}
\usepackage{dcolumn}

\usepackage{natbib}

\DeclareMathAlphabet{\mathsc}{OT1}{cmr}{m}{sc}
\def\testbx{bx}%
\DeclareRobustCommand{\ion}[2]{%
\relax\ifmmode
\ifx\testbx\f@series
{\mathbf{#1\,\mathsc{#2}}}\else
{\mathrm{#1\,\mathsc{#2}}}\fi
\else\textup{#1\,{\mdseries\textsc{#2}}}%
\fi}

\newcommand{\HI}{\ion{H}{i}}
\newcommand{\CI}{\ion{C}{i}}

\newcommand{\SiII}{\ion{Si}{ii}}

\newcommand{\NiII}{\ion{Ni}{ii}}

\newcommand{\FeII}{\ion{Fe}{ii}}
\newcommand{\MgI}{\ion{Mg}{i}}
\newcommand{\MgII}{\ion{Mg}{ii}}
\newcommand{\MnII}{\ion{Mn}{ii}}
\newcommand{\ZnII}{\ion{Zn}{ii}}

\newcommand{\CrII}{\ion{Cr}{ii}}

\def\kms{km\,s$^{-1}$}
\def\ms{m\,s$^{-1}$}

\def\zabs{$z_{\rm abs}$}
\def\zem{$z_{\rm em}$~}
\def\dxx{$\Delta x/x$}
\def\daa {$\Delta \alpha/\alpha$}
\def\dmm{$\Delta \mu/\mu$}

\begin{document}

\title[constraining  fundamental constants at $z \sim 1.3$]{Constraining the variation of
fundamental constants at $z\sim1.3$ using 21-cm absorbers\thanks{Based on data obtained with 
UVES at the Very Large Telescope of the European Southern Observatory 
(Prgm. ID 082.A-0569A, 085.A-0258A, 66.A-0624A, 68.A-0600A, 072.A-0346A, and 074.B-0358A)}}
\author[Rahmani, H.]{H. Rahmani$^{1}$, R. Srianand$^{1}$, N. Gupta$^{2}$, P. Petitjean$^{3}$, P. Noterdaeme$^{3}$ and
\newauthor D.  Albornoz V\'{a}squez$^{3}$\\
$^{1}$ Inter-University Centre for Astronomy and Astrophysics, Post Bag 4,  Ganeshkhind, Pune 411\,007, India \\
$^{2}$ Netherlands Institute for Radio Astronomy (ASTRON), Postbus 2, 7990 AA, Dwingeloo, The Netherlands \\
$^{3}$ Universit\'e Pierre et Marie Curie, CNRS-UMR7095, Institut d'Astrophysique de Paris, 98bis Boulevard Arago, 75014 Paris, France \\
}
\date{Accepted. Received; in original form }
\pagerange{\pageref{firstpage}--\pageref{lastpage}} \pubyear{2012}
\maketitle
\label{firstpage}

\begin {abstract}
{ 
We present high resolution optical spectra obtained with the Ultraviolet and Visual
Echelle Spectrograph  (UVES) at the Very Large 
Telescope (VLT) and 21-cm absorption spectra obtained with the Giant Metrewave Radio Telescope (GMRT) 
and the Green Bank Telescope (GBT) of five quasars along the line of sight of  which 21-cm absorption systems 
at $1.17 ~< ~z ~< ~1.56$ have been detected previously. We also present milliarcsec scale radio images
of these quasars obtained with the Very Large Baseline Array (VLBA). 
We use the data on four of these systems to constrain the time variation of 
$x\equiv g_{\rm p}\alpha^2/\mu$ where $g_{\rm p}$ is the proton gyromagnetic factor, 
$\alpha$ is the fine structure constant, and $\mu$ is the  proton-to-electron 
mass ratio. We carefully evaluate the systematic uncertainties in redshift measurements
using cross-correlation analysis and repeated Voigt profile fitting.
In two cases we also confirm our results by analysing optical spectra obtained with the Keck telescope.
We find the weighted and the simple means of \dxx\ to be respectively 
$-(0.1 \pm1.3)\times 10^{-6}$ and $ (0.0 \pm1.5)\times 10^{-6}$ 
at the mean redshift of $<z> = 1.36$ corresponding to a look back 
time of $\sim$9~Gyr. This is the most stringent constraint ever obtained on \dxx. 
If we only use the two systems towards quasars unresolved at milliarcsec scales,
we get the simple mean of \dxx~=~$+(0.2 \pm1.6)\times 10^{-6}$. 
Assuming constancy of other constants we get  \daa~=~$(0.0 \pm0.8)\times 10^{-6}$ which is a factor of 
two better than the best constraints obtained so far using the  Many Multiplet Method.
On the other hand assuming $\alpha$ and $g_{\rm p}$ have not varied we derive  
\dmm\ = $ (0.0 \pm1.5)\times 10^{-6}$ which is again the best limit ever obtained on the variation of $\mu$ 
over this redshift range. Using independent constraints on \daa\ at $z < 1.8$ and \dmm\  at $z\sim 0.7$ available in the
literature we get $\Delta g_{\rm p}/g_{\rm p} \le 3.5 \times10^{-6} (1\sigma)$. 
} 
\end{abstract}

%
%
%
%
%
%
%
%
\section{Introduction}
Most of the successful physical theories rely on the constancy of
 a few fundamental quantities such as the fine structure constant, $\alpha=e^2/\hbar c$,  
the proton-to-electron mass ratio, $\mu$, etc.
Some modern theories of high energy physics that try to unify 
the fundamental interactions predict the variation of these dimensionless fundamental 
constants over cosmological scales \citep[see]
[and references therein for more details]{Uzan03}. 
Current laboratory constraints exclude any significant variation of 
these constants over solar system scales and  on geological 
time scales \citep[see][]{Olive04,Petrov06,Rosenband08}.
It is not observationally/experimentally excluded however that they 
could vary over cosmological scales. 
Therefore, constraining time and spatial variations of fundamental constants of physics 
will have a great impact on understanding the true behavior of  the nature. 

\citet{Savedoff56} first pointed out the
possibility of using redshifted atomic lines from distant objects to test the
evolution of dimensionless physical constants. 
Initial  attempts in this field mainly used the 
Alkali-Doublet (AD) method to constrain $\alpha$ variation 
\citep{Savedoff56,Bahcall67b,Wolfe76,Levshakov94,Varshalovich96,Cowie95,Varshalovich00,Murphy01, Chand05}.  
While the AD method is simple, and  least affected by systematics related 
to ionization and chemical inhomogeneities, 
the limits achieved on $\Delta\alpha/\alpha$ $\equiv(\alpha_{\rm z}-\alpha_{0})/\alpha_{0}$  
are not usually stringent \footnote{Here 
$\alpha_{\rm z}$  and $\alpha_0$ are the measured values of $\alpha$ at any redshift, $z$, and in the laboratory 
on Earth}. The most precise value reported to date using this method
being $\Delta\alpha/\alpha =-(0.02\pm0.55)\times10^{-5}$ \citep{Chand05} over a redshift range 
of 1.59$\le z\le$ 2.92.

\citet{Dzuba99a,Dzuba99b} and \citet{Webb99} introduced 
the Many-Multiplet (MM) method as a generalization of the AD method, 
in which one correlates different multiplets from different ions 
simultaneously.  Applying this method on a sample of 128 absorbers observed 
at high spectral resolution with the Keck telescope, \citet{Murphy03} claimed a detection,
$\Delta\alpha$/$\alpha$=$-(0.57\pm0.10)\times10^{-5}$,  
over the redshift range 0.2$\le z \le$3.7.
However, this result was not confirmed by \citet{Srianand04} and \citet{Chand04}
who used higher  signal-to-noise ratio (SNR~$\sim$~70 per pixel), high spectral 
resolution ($R\ge$45000) UVES/VLT  data of 23 \MgII\ systems 
detected towards 18 quasars in the redshift range $0.4\le z\le 2.3$  and found 
${ \Delta\alpha/\alpha}$~=~${ (-0.06\pm0.06)\times10^{-5}}$.
This analysis was criticized by \citet{Murphy07prl}. However,  from 
the reanalysis of the UVES data, using the Voigt profile fitting code
VPFIT\footnote{http://www.ast.cam.ac.uk/~rfc/vpfit.html}, \citet{Srianand07b} confirmed the null result albeit with
 larger error bars (i.e ${ \Delta\alpha/\alpha}$~=~${ (0.01\pm0.15)\times10^{-5}}$). 
Other analysis using only
\FeII\ transitions in  two particularly well suited absorption systems at $z = 1.15$ and 
$z = 1.84$ failed to confirm any variation in $\alpha$ 
\citep{Quast04, Levshakov06, Chand06, Levshakov07}. 
Recently, \citet{Webb11} have reported the  results of the analysis of 153 systems present in 
quasar spectra observed with VLT/UVES. 
They find that $\alpha$ increases with increasing cosmological distance from the Earth.  
Moreover for $z<1.8$, they confirm the results by \citet{Srianand07b},
${\Delta\alpha/\alpha}$~=~$-(0.6\pm1.6)\times10^{-6}$. 
However combining their new VLT measurements with their previous Keck measurements they
suggest the possibility for  a spatial variation of $\alpha$ and
speculate on the existence of an $\alpha-$dipole. 
If true, this dipole is very difficult to explain theoretically \citep{Olive11}.
While the MM-method provides improved precision, it is affected by systematics 
related to ionization, chemical inhomogeneities and isotopic composition. 
The effects of inhomogeneities can be canceled using a large sample of absorption 
systems but the effects of isotopic composition will likely to remain an issue.

Most of the existing theories predict that the proton-to-electron mass ratio $\mu$ 
should vary much more than $\alpha$ \citep[for example see][and references therein]{Olive02,Dent03,Dine03} 
though some predicts the reverse \citep[][]{Dent08}. 
The variations of $\mu$ can be probed using 
H$_2$ Lyman and Werner band absorption lines \citep{Varshalovich93}. H$_2$ molecules are occasionally
detected in high redshift damped Lyman-$\alpha$ systems 
\citep{Petitjean00,ledoux03,noterdaeme08,Srianand12} with only a handful of them being 
suitable for probing the variation of $\mu$. No clear indication of any variation in $\mu$ in excess of 1 part in
10$^5$ is seen in the existing data for  $z\ge2$ \citep{Ivanchik05, Reinhold06, King08,Thompson09b, Wendt11, vanWeerdenburg11}. 
By comparing inversion line transitions of NH$_3$ with the rotational transitions of other molecules, a strong
constraint on $\Delta \mu/\mu$ can be obtained \citep{Murphy08}. 
At present such an exercise is possible for only two gravitationally lensed systems at $z<1$ 
\citep{Henkel05, Henkel08}. The best reported constraint is 
$\Delta \mu/\mu \le 3.6\times 10^{-7} (3\sigma)$ at $z$ = 0.685 by \citet{Kanekar11}.
Detecting  more NH$_3$ absorption towards normal quasars is required to reduce 
systematics related to the usage of lensed quasars (See \citet{Henkel08} for discussions on various other systematics). 

As the energy of the 21-cm transition is proportional to, $x\equiv\alpha^2 g_{\rm p}/ \mu$, high resolution optical spectra 
and 21-cm spectra can be used together to probe the combined variation of these constants \citep{Wolfe76}.  
Constraints of the order of $\sigma (\Delta x/x)  \lesssim 10^{-5}$ were obtained towards individual systems \citep{Cowie95,Kanekar06,Srianand10}. 
\citet{Tzanavaris07} derived ${\Delta x/ x} = (0.63\pm0.99)\times 10^{-5}$ for 
a sample of nine 21-cm absorbers with $0.23~ <~ z~ <~ 2.35$. The majority of the 21-cm spectra used in this study  were digitally scanned 
from the  printed literature and the UV-optical data were obtained mainly with VLT/UVES. 
Better constraints can be derived from  higher quality spectra in the radio and optical wavelength ranges  
of a well selected sample of 21-cm absorbers.
This is possible now thanks to systematic surveys for 21-cm absorption
towards strong Mg~{\sc ii} absorbers \citep[e.g.][]{Gupta09}. This work has 
resulted in the detection of 9 new 21-cm  absorption systems over a 
narrow redshift range (i.e 1.05$\le z\le$1.45) that can be used for constraining $\Delta x/x$.

While  this technique is very powerful there are two 
issues that introduce systematic uncertainties in the measurements. 
These are: (i) the identification of the optical component corresponding to the 
gas that produces the 21-cm absorption and (ii) the fact that the radio and optical 
sources could probe different volumes of the absorbing gas as the radio  
emitting region in quasars is in general extended compared to the UV emitting region. 
It has been suggested that 
the gas detected by their \CI\ and/or $\rm H_2$ 
absorption is closely associated with the 21-cm gas 
\citep[][]{Cowie95,Srianand10}. 
However, only few 21-cm absorbers show detectable \CI\  and $\rm H_2$ absorption 
and even in these cases velocity 
offsets up to 1-2 \kms\ are noticed \citep[][]{Srianand12}. 
All these indicate \CI/$\rm H_2$ and 21-cm absorption need not originate from the same physical region. 
Another option is to connect 21-cm absorption to absorption 
from singly ionized species that trace \HI\ gas.   
For example \citet{Tzanavaris07} have associated  
the pixel with strongest absorption in the UV
with the pixel with the strongest 21-cm absorption. As neighboring 
pixels are correlated in optical spectra, the redshift of the strongest 
metal absorption component will be better defined by using simultaneous Voigt profile 
fits to the absorption lines. This is the method we adopt in the analysis presented here. 
The second uncertainty discussed above can be minimized by 
selecting absorbers towards quasars that are compact at milliarcsec scales. While 
individual measurements may  not be completely free of these systematics, even after 
careful consideration of the specific properties of the system,
it should be possible to minimize them and get a statistically reliable measurement using large 
sample of absorbers. 

As different methods used for constraining the fundamental constants suffer from 
different systematic effects it is important to increase 
the number of measurements based on each method to 
address the time and space variation of different constants. Here we provide 
new measurements of \dxx\ using a new sample of 21-cm absorbers.

We have selected 5 systems from the literature [4 from \citet{Gupta09}
and one from \citet{Kanekar09}] previously known to be associated with narrow 21-cm absorption lines
towards radio sources that are compact at arcsecond scales. We have obtained
high resolution UV and radio data of the quasars together with 
high resolution Very Large Baseline Array (VLBA) images. We report here the analysis of
this dataset. This paper
is organized as following. In Section 2, we present details of optical
and radio observations and data reduction. In Sections 3 and 4 we provide
details of Gaussian fits to the 21-cm absorption lines and Voigt profile
fitting of the UV lines. In Section 5, we summarize our $\Delta x/x$ 
measurements in individual systems and discuss the associated systematic 
errors. In Section 6 we discuss the results and conclude. 
 We use simultaneous Voigt profile 
fits to identify the redshift of the strongest UV component closest to the 21-cm 
absorption. We also discuss the results if we adopt the method used by 
 \citet{Tzanavaris07}.
\begin{table*}
\caption{Log of the optical spectroscopic observation with VLT/UVES}
\begin{center}
\begin{tabular}{cccccccc}
\hline
\hline
Source name & Exposure name &Date & Starting   & Exposure    & Setting  & Seeing & Airmass\\
            &               &     & Time (UT)  &  (sec)      &          &(arcsec)&  \\
~~~~~~~~(1)&  (2)  &   (3) &  (4)  & (5)  & (6) & (7) & (8)\\
\hline
J0108$-$0037& EXP1 &2008-11-21    &01:15:02     & 3700  & 390+580 &0.64 &1.11 \\
            & EXP2 &2008-11-23    &02:41:45     & 3700  & 390+580 &0.77 &1.13 \\
            & EXP3 &2008-11-25    &01:52:08     & 3700  & 390+580 &0.89 &1.10 \\
            & EXP4 &2008-12-03    &01:11:20     & 3690  & 390+580 &0.71 &1.10 \\
J1623+0718  & EXP1 &2010-05-08    &05:48:49     & 3340  & 390+580 &1.33 &1.18 \\
            & EXP2 &2010-08-07    &01:08:41     & 3340  & 390+580 &0.82 &1.23 \\
            & EXP3 &2010-08-08    &00:45:37     & 3340  & 390+580 &0.86 &1.20 \\
            & EXP4 &2010-08-09    &00:56:50     & 3340  & 390+580 &0.70 &1.22 \\
J2340$-$0053& EXP1 &2008-10-02    &02:39:30     & 4500  & 390+580 &0.88 &1.13 \\
            & EXP2 &2008-10-05    &02:01:02     & 4500  & 390+580 &0.82 &1.17 \\
            & EXP3 &2008-10-05    &03:25:10     & 4500  & 390+580 &0.88 &1.09 \\
            & EXP4 &2008-10-06    &00:28:59     & 4500  & 390+580 &0.90 &1.50 \\
            & EXP5 &2008-10-06    &01:55:32     & 4500  & 390+580 &0.79 &1.18 \\
            & EXP6 &2008-10-28    &04:41:03     & 4500  & 390+580 &0.79 &1.45 \\
J2358$-$1020& EXP1 &2010-08-04    &06:27:07     & 3340  & 390+580 &0.75 &1.10 \\  
            & EXP2 &2010-08-06    &04:43:45     & 3340  & 390+580 &0.69 &1.40 \\  
            & EXP3 &2010-08-06    &05:50:01     & 3340  & 390+580 &0.71 &1.16 \\  
            & EXP4 &2010-08-06    &06:55:07     & 3340  & 390+580 &0.64 &1.05 \\  
            & EXP5 &2010-08-06    &08:00:14     & 3340  & 390+580 &0.65 &1.04 \\  
            & EXP6 &2010-08-06    &09:05:21     & 3340  & 390+580 &0.77 &1.10 \\  
            & EXP7 &2010-08-07    &04:53:34     & 3340  & 390+580 &0.78 &1.34 \\  
J0501$-$0159& EXP1 &2000-10-21    &06:13:17     & 3600  & 437+750 &0.61 &1.17 \\  
            & EXP2 &2000-10-23    &04:08:46     & 3600  & 346+580 &0.53 &1.73 \\  
            & EXP3 &2001-10-16    &07:20:51     & 5400  & 346+570 &0.46 &1.10 \\  
            & EXP4 &2004-10-21    &04:38:08     & 4500  & 390+564 &0.63 &1.25 \\  
            & EXP5 &2004-10-21    &05:42:38     & 5400  & 390+564 &0.79 &1.56 \\  
            & EXP6 &2004-10-21    &07:05:51     & 3600  & 437+860 &1.17 &1.10 \\  
            & EXP7 &2004-10-22    &04:38:08     & 3600  & 437+860 &0.84 &1.29 \\  
            & EXP8 &2004-10-22    &05:42:42     & 4500  & 390+564 &0.89 &1.12 \\  
            & EXP9 &2004-10-22    &07:05:55     & 4500  & 390+564 &0.59 &1.81 \\  
            & EXP10&2004-10-22    &08:04:58     & 3360  & 437+860 &1.00 &1.09 \\  
\hline
\end{tabular}
\end{center}
\begin{flushleft}
Column 1: Source name; Column 2: Assigned name for the exposure.  
Column 3: Date of observation; Column 4: Starting time of exposure; Column 5: Exposure time;
Column 6: Spectrograph settings; Column 7: Seeing in arcsec; Column 8: Airmass at the beginning 
of the exposures.
\end{flushleft}
\label{MgIIslist}
\end{table*}
%
\section{Observations and data reduction}
\subsection{Optical spectroscopy}\label{MgIIsample_optical}
The optical spectroscopic observations of quasars  
were carried out with UVES  \citep[][]{Dekker00} at the VLT UT2 8.2-m telescope  at Paranal (Chile) 
in service mode [Programs 082.A-0569A and 085.A-0258A].  
All observations  were performed using the standard beam splitter with the 
dichroic \#2 (setting 390+580) that covers roughly  from 330 to 450 nm 
on the BLUE CCD and from 465 to 578 nm and 583 to 680 nm on the two RED CCDs. 
Slit width of 1 arcsec and CCD readout with 2x2 binning were used for all 
the observations  resulting in a pixel size of $\approx$ 1.7 \kms\ and 
spectral resolution of $\approx$45000. \citet{D'Odorico00} have shown
that the resetting of the grating between an object exposure and the
ThAr calibration lamp exposure can result in an error of the order of a few hundred meters per second 
in the wavelength calibration. To minimize this effect each science exposure was
followed immediately by an attached set of 5 ThAr lamp exposures. 
In the case of J0501$-$0159 we have retrieved all the UVES data available in the
ESO archive. As these spectra were not acquired specifically for 
constraining the variation of fundamental constants there is 
no attached calibration lamp exposure taken along with the science observations. 
However this data were reduced using the available lamp spectra closest in time. 

The data were reduced with the UVES Common Pipeline Library (CPL) data reduction 
pipeline release 4.7.8\footnote{http://www.eso.org/sci/facilities/paranal/instruments/uves/doc/} 
using the optimal extraction method. We used  $\rm 4^{th}$ order polynomials to find the 
dispersion solution. The number of suitable ThAr lines used for wavelength calibration 
was always larger than 400 and the rms error  was found to be in the range 70 -- 80  \ms\ with zero average.
However, this error applies only to
regions very close to the ThAr emission lines that are used to compute the wavelength solution. 
In principle the calibration error in the regions in between ThAr emission line 
can be typically of the order of a few hundred meters per second \citep[for example see][]{Agafonova11}.  

All the spectra were corrected for the motion of the observatory 
around the barycenter of the Sun-Earth system. The velocity component of the 
observatory's barycentric motion towards the line of sight to the object was 
calculated at the exposure mid point  (see Table \ref{MgIIslist}). Conversion of air to vacuum wavelengths  
was performed using the formula given in \citet{Edlen96}. 
For the co-addition of the different exposures, we interpolated the individual spectra and their errors to a common wavelength 
array and then computed the weighted mean using weights estimated from the errors in each pixel. 
In order to fit the continuum we 
considered only specific regions (20-100 \AA) around the absorption lines of interest and fitted the points without any absorption 
with a lower order cubic spline. 

Voigt profile fits of the absorptions from different species have been performed using VPFIT, 
version 9.5.  While simultaneously fitting absorption profiles of a 
system, we assumed  that all the singly ionized species 
(e.g. \FeII, \SiII, \ZnII, etc.) are  kinematically associated with the 
same gas.  We also assumed the velocity broadening is predominantly 
turbulent. 
Therefore, we used the same $z$ and $b$ parameters for a given
component for all the species. 
The error on the redshift of individual Voigt profile components depends on the 
statistical error from the fitting procedure and the systematic errors related to 
the procedure itself and the wavelength calibration. 
The VPFIT program estimates errors using only the diagonal terms of the covariance matrix. 
Although the reliability of the errors have been confirmed for unblended components
\citep[see][]{King09,Carswell11}, 
errors from VPFIT are underestimated in the case of blended components. To account for this and 
other systematic errors discussed above, we perform Voigt profile fits for a given system several times 
(see Section \ref{Voigt_r} for details).
In Table \ref{tabatomic} we summarize the laboratory wavelengths and
oscillator strengths of all transitions that are used in this study. 
In this table we also give a short name (id) to specific transitions 
for future reference in the text.

\begin{table}
\caption{Adopted atomic data for different species used in this study}
\begin{center}
\begin{tabular}{lcllc}
\hline
\hline
Species & id &wavelength       &Ref &   oscillator$\dagger$          \\
        & &~~~~~~     (\AA)  &  &strength                   \\
\hline
\SiII  & a  &  1808.01288     & [2]  & 0.00208    \\
\CrII  & b1 &  2056.25682     & [2]  & 0.1030     \\
\CrII  & b2 &  2062.23594     & [2]  & 0.0759     \\
\CrII  & b3 &  2066.16391     & [2]  & 0.0512     \\
\MnII  & c1 &  2576.87534     & [2]  & 0.361      \\
\MnII  & c2 &  2594.49669     & [2]  & 0.280      \\
\MnII  & c3 &  2606.45883     & [2]  & 0.198      \\
\FeII  & d1 &  1608.45081     & [3]  & 0.0577     \\
\FeII  & d2 &  1611.2005      & [1]  & 0.00138    \\
\FeII  & d3 &  2249.8768      & [1]  & 0.00182    \\
\FeII  & d4 &  2260.77934     & [2]  & 0.00244    \\
\FeII  & d5 &  2344.21282     & [2]  & 0.114      \\
\FeII  & d6 &  2374.46013     & [2]  & 0.0313     \\
\FeII  & d7 &  2382.76411     & [2]  & 0.320      \\
\FeII  & d8 &  2586.64937     & [2]  & 0.0691     \\
\FeII  & d9 &  2600.17223     & [2]  & 0.239      \\
\NiII  & e1 &  1454.842       & [1]  & 0.0276     \\
\NiII  & e2 &  1467.259       & [1]  & 0.0063     \\
\NiII  & e3 &  1467.756       & [1]  & 0.0099     \\
\NiII  & e4 &  1502.148       & [1]  & 0.006      \\
\NiII  & e5 &  1703.4119      & [1]  & 0.006      \\
\NiII  & e6 &  1709.6042      & [1]  & 0.0324     \\
\NiII  & e7 &  1741.5531      & [1]  & 0.0427     \\
\NiII  & e8 &  1751.9157      & [1]  & 0.0277     \\
\ZnII  & f1 &  2026.13695     & [2]  & 0.501      \\
\ZnII  & f2 &  2062.66028     & [2]  & 0.246      \\
\CI    & g1 &  1560.3092      & [1]  & 0.0774     \\
\CI    & g2 &  1656.9284      & [1]  & 0.149      \\
\MgI   & h1 &  2026.47504     & [2]  & 0.113      \\
\MgI   & h2 &  2852.96282     & [2]  & 1.83       \\
\hline
\end{tabular}
\end{center}
\begin{flushleft}
 References. [1] -- \citet{Morton03}; [2] -- \citet{Aldenius09}; [3] -- \citet{Nave11}\\
$\dagger$ Oscillator strengths are from \citet{Morton03} apart from the \NiII$\lambda$1454 
where we use the value from \citet{Zsargo98} that provide best fit to our data while 
being consistent within error to the values given in \citet{Morton03}. 
\end{flushleft}
\label{tabatomic}
\end{table}
%
%
%

\subsubsection{Systematic errors in wavelength calibration} 
\label{sys_error_opt}

The shortcomings of the ThAr wavelength calibration of quasars spectra taken 
with VLT/UVES has already  been discussed by number of authors 
\citep{Chand06,Levshakov06,Molaro08,Thompson09b,Whitmore10,Agafonova11}. 
\begin{figure} 
\centering
\includegraphics[width=0.8\hsize,bb=18 18 594 774,clip=,angle=90]{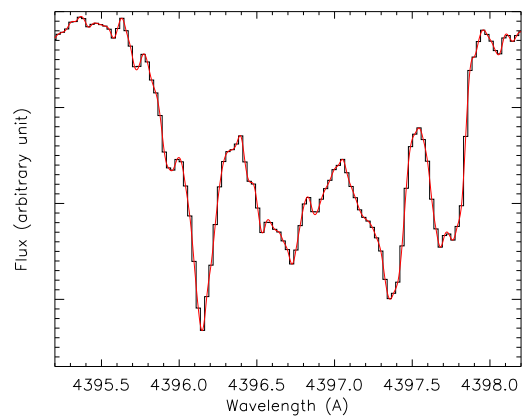}
\caption{Example of the interpolation of a spectrum using Neville's algorithm. 
The histogram curve is an absorption profile and the smooth curve gives   
the interpolated spectrum achieved using 
Neville's algorithm.} 
\label{Interp_examp}
\end{figure}
\begin{figure} 
\centering
\includegraphics[width=0.8\hsize,bb=18 18 594 774,clip=,angle=90]{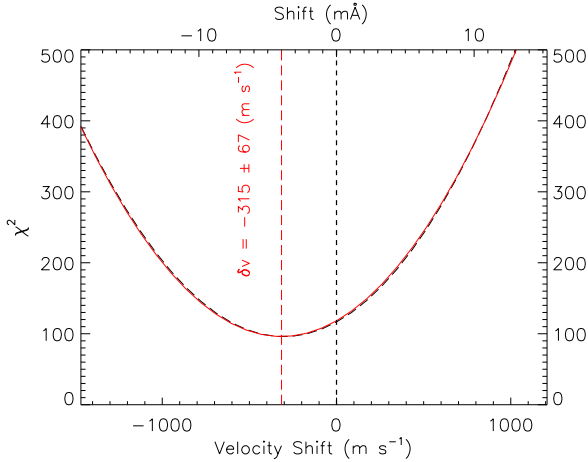}
\caption{Velocity shift between the combined spectra 
and one of the exposures for the \FeII\ $\lambda$1608
absorption profile in J2340$-$0057. The dashed curve shows the $\chi^2$ 
(see \ref{sys_error_opt}) as a function of the applied velocity 
shift and the continuous curve is the best fitted parabola. The short and long dashed 
vertical lines respectively mark the 
zero velocity and velocity of $-$315~\ms\ at which the $\chi^2$ is 
minimum.} 
\label{Vshift_examp}
\end{figure}
Overall velocity shifts of the order of a few hundred meters per second have been 
observed between the spectra of the same object calibrated with an iodine 
cell spectrum or with a ThAr calibration lamp \citep{Whitmore10}. 
The latter authors have also shown that  intra-order 
velocity shifts of more than 200 \ms\  are present within a given exposure.
Therefore, while we have taken enough care during the data reduction, 
wavelength uncertainties due to these systematics still remain.
We therefore performed several tests to estimate 
these systematic effects.

\subsubsection{Cross-correlation analysis}\label{cross_cor_ana}
We cross-correlate  individual spectra (in a window comprising of the 
absorption lines associated with the systems of interest) 
with the combined spectrum to estimate the velocity offset between them. 
For this we first rebin each pixel ($\delta v$$\sim$2.0 \kms) 
to 25 sub-pixels with a velocity width of $\sim$80 \ms.  This rebinning is 
done by interpolating the spectrum with a polynomial using Neville algorithm 
(see Fig. \ref{Interp_examp}). 
For finding the relative shift between 
the combined spectrum $I$ and the spectrum $I_{\rm i}$ of the i'th exposure,
we proceed as follows: we fix $I$ and 
 shift $I_{\rm i}$ relative to  $I$ by steps of 0.04 times the
original pixel size (i.e 80 \ms). At each step characterized by the shift $\delta \lambda$, 
a $\chi^2$ is calculated from
the  difference in fluxes between $I$ and $I_{\rm i}$ and the flux error of $I_{\rm i}$: 
$\chi^2 (\delta \lambda) = \Sigma_{\rm j} 
[I_{\rm j} - I_{\rm i,j}(\delta \lambda)]^{2}/\sigma_{\rm i,j}^2(\delta \lambda)$, where  $I_{\rm i,j}$ is 
the normalized flux of the j'th pixel of the i'th exposure 
with the error $\sigma_{\rm i,j}$.
This $\chi^2$ is a function of $\delta \lambda$ and is minimum when the two profiles are 
aligned. We fit the function $\chi^2(\delta \lambda)$ with a parabola and the value 
of $\delta \lambda_{\rm min}$ at which  the $\chi^2$ is minimum is taken as the wavelength offset between the two 
absorption profiles (See Fig. \ref{Vshift_examp}). Following the standard statistical procedure 
we assign 1$\sigma$ errors to this $\delta \lambda_{\rm min}$ by computing the required change in $\delta \lambda$ so that 
$\Delta \chi^2 = \chi^2 - \chi^2_{\rm min} = 1$. 
This procedure is similar to that implemented by \citet{Agafonova11} to find the shift between their 
spectra  but with the difference 
that they use the simple sum of the square of the differences in fluxes instead of $\chi^2$. 
Using $\chi^2$ has the advantage that we can associate an error to the measured shift. 
For each exposure we measure the shifts (of all transitions used in our Voigt profile fitting)  
along with the errors. 
Having measured a shift and error for each  transition in each exposure we find their weighted 
mean and weighted standard deviation as an estimate of the systematic 
error  due to a constant shift in the redshift of the absorbing system. 
In addition, the correlation analysis not only allows us to 
identify exposures with abnormally large wavelength shifts but also to
identify absorption lines that may be affected by calibration uncertainties in one of 
the exposures. Results of this exercise for  four  quasars in our sample are summarized 
in Tables \ref{tab0108shift_opt}, \ref{tab1623shift_opt}, \ref{tab2340shift_opt}, and 
\ref{tab2358shift_opt} of the Appendix.  
It is clear from these tables that individual exposures 
have typical (rms) shifts of up to $\sim$350 \ms.
\subsubsection{Repeated Voigt profile fitting analysis}\label{Voigt_r}
We use simultaneous fitting of several transitions to measure the redshift of a given component 
and the associated error. 
To estimate the latter we perform repeated Voigt 
profile fitting using several combinations of spectra excluding one exposure at
a time. This exercise allows us to understand the influence of
individual exposures on our final redshift measurement. Similarly
using the final combined spectra we perform repeated Voigt profile
fitting including and excluding different transitions. This will
allow us to estimate the redshift uncertainties due to the choice 
of lines used in the Voigt profile fitting and also the random intra-order shifts. 

It is known that the wavelength calibration is most accurate in regions close to 
the ThAr emission lines that are used to derive the pixel to wavelength solution  \citep[see][]{Agafonova11}. 
Therefore, we performed Voigt profile fitting of those transitions that have 
at least one ThAr emission line (that was used for wavelength calibration) 
within $\pm50$ \kms\ from the UV-optical component  that coincides with 
the 21-cm component. 

We use the results of the above exercises  along with 
the results of the cross-correlation analysis to quantify the final  errors  
in the redshift measurements.
\subsection{Archival Keck/HIRES spectrum}
For two quasars studied here (J2340$-$0053 and J0501$-$0159) high resolution
echelle spectra were obtained with Keck/HIRES by Prof. Prochaska and collaborators as part 
of their database archive for abundance studies in DLAs \citep{Prochaska01_HIRES,Prochaska07}.
The wavelength coverage  in both cases is less than that of our VLT/UVES spectra. 
The spectral resolution of the J2340$-$0053  Keck/HIRES spectrum 
is roughly the same as our VLT/UVES spectrum (i.e 6.0 \kms) but its SNR is less than ours. 
In the case of J0501$-$0159 the spectral resolution of Keck/HIRES is $\sim$ 8.0 \kms\ and both 
spectra have comparable SNR. We fit the absorption 
profiles in the Keck/HIRES in order to  compare the results obtained with the
two telescopes.  
This exercise helps us to understand the systematic errors in the wavelength calibration 
and especially the existence of any global shift between the two spectra.

\subsection{GMRT observations and morphology of the background sources} \label{MgIIsample_radio}
\citet{Gupta09} used a bandwidth of 1\,MHz split into 128 frequency channels in the course of their 
GMRT survey  for 21-cm absorption in strong \MgII\ absorbers. This yields
a velocity resolution of $\sim$4 \kms\  per channel. In our new GMRT observations
for three sources (J0108$-$0038, J2340$-$0053, and J2385$-$1020)
we have used  a band width of 0.25\,MHz split into 128 channels yielding
a channel resolution of $\sim$1 \kms. To increase the spectral SNR, each object was observed
for 24 to 30 hrs (i.e in three full synthesis observations). In the
case of J1623$+$0718  a bandwidth of 0.5 MHz was used to adequately cover both the 
21-cm components. Spectral resolution in this case is  $\sim2$ \kms.

The data were acquired in the two orthogonal polarization channels RR and LL.  
For the flux density/bandpass calibration of GMRT data, standard flux density 
calibrators were observed for 10-15\,min every two hours. A phase calibrator 
was also observed for 10\,min every 45\,min to get reliable phase solutions.    
The GMRT data were reduced using the NRAO AIPS package following the standard 
procedures. Special care was taken to exclude the baselines and time stamps affected 
by the radio frequency interference (RFI). 
The spectra at the quasar positions were extracted from the RR and LL 
spectral cubes and compared for consistency. If necessary, a first-order 
cubic-spline was fitted to remove the residual continuum.  
The two polarization channels were then combined to get the stokes I spectrum 
which was then shifted to the heliocentric frame. We used the AIPS
task CVEL to correct the observed data for the Earth's motion and
rotation. 
We obtained the mean spectrum weighting the flux by
 the inverse rms square in the line free channels.

\subsubsection{Redshift uncertainties}
\begin{table}
\caption{Shifts (in units of \ms) between individual 21-cm spectra  relative to the combined one 
for J0108$-$0037 at the position of 21-cm absorption}
\begin{center}
\begin{tabular}{cccccc}
\hline
\hline
$\rm LL_1$  &$\rm RR_1$  & $\rm LL_2$  &$\rm RR_2$  &$\rm LL_3$  &$\rm RR_3$   \\ [1.0ex]
\hline
 47$\pm$55     &99$\pm$96      & 71$\pm$104     &-180$\pm$133    & -178$\pm$191 & -249$\pm$192 \\
\hline
   \multicolumn{3}{l}{weighted mean}                     &   +17     &        &       \\
   \multicolumn{3}{l}{weighted standard deviation}       &   122     &        &       \\
\hline
\end{tabular}
\end{center}
\begin{flushleft}
\end{flushleft}
\label{tabshift_21}
\end{table}
We can use the cross-correlation analysis described in Section \ref{cross_cor_ana}
to search for any possible frequency off-set between spectra of the same object
obtained at different epochs and through different polarization channels. 
To avoid the effect of poor SNR, we consider here only J0108$-$0037.
We have useful data obtained during three epochs and hence 6 individual spectra to carry out a  
correlation analysis of individual spectra relative to the final combined one. 
The results are summarized in Table~\ref{tabshift_21}.
Here $\rm LL_i$ and $\rm RR_i$ are corresponding to the LL and 
RR polarizations of the i'th observation of this source. 
The weighted standard deviation of these observed shifts is 122 \ms\ 
and observed values are up to 249 \ms. 
We will thus consider the systematic error in radio frequency calibration 
to be 122 \ms\ or equivalently 0.4$\times 10^{-6}$ in $\Delta x/x$. 

Below,  we will give the results of Gaussian fitting of the 21-cm lines 
using the combined spectrum and spectra obtained through 
individual polarization channels. 
The statistical errors in 21-cm redshift determination are found to be much 
larger than the above quoted systematic shift.

\subsubsection{Milliarcsec images of the background sources}
%
\begin{figure*}
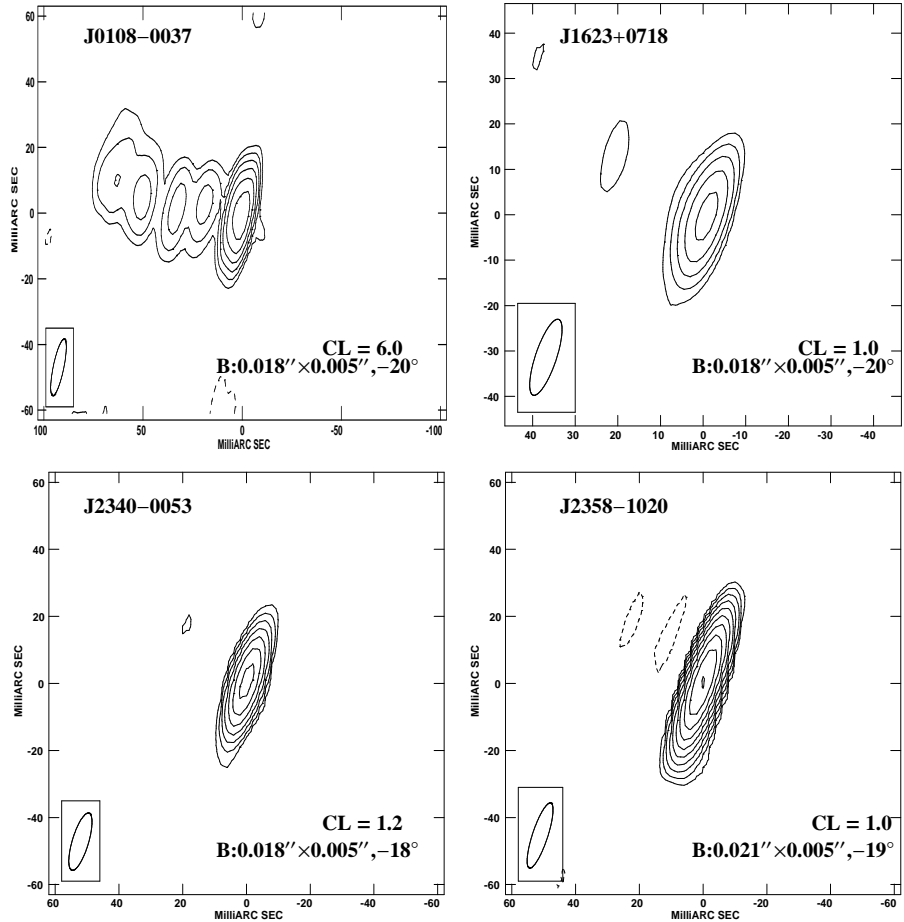

\begin{center} 
\centerline{
\vbox{
\hbox{
\includegraphics[height=6.0cm,width=6.2cm,angle=-90]{J0108MAP_NOLABELS.PS}
\includegraphics[height=6.0cm,width=6.2cm,angle=-90]{J1623MAP_NOLABELS.PS}
}
\hbox{
\includegraphics[height=6.0cm,width=6.2cm,angle=-90]{J2340MAP_NOLABELS.PS}
\includegraphics[height=6.0cm,width=6.2cm,angle=-90]{J2358MAP_NOLABELS.PS}
}
}
}
\end{center}
\caption{Contour plots of VLBA images at 1.4\,GHz.   The restoring beam, shown as the ellipse, and the first 
contour level (CL) in mJy\,beam$^{-1}$ are provided at the bottom of each image.  
The contour levels are plotted as CL$\times$($-$1, 1, 2, 4, 8,...)\,mJy\,beam$^{-1}$.
}
\vskip -14.0cm
\begin{picture}(400,400)(0,0)

\put( 060,378){\bf J0108$-$0037}
\put( 150,260){\bf       CL = 6.0}
\put( 110,253){\bf       B:0.018$^{\prime\prime}$$\times$0.005$^{\prime\prime}$,$-$20$^\circ$ }
\put( 240,378){\bf J1623$+$0718}
\put( 330,260){\bf       CL = 1.0}
\put( 290,253){\bf       B:0.018$^{\prime\prime}$$\times$0.005$^{\prime\prime}$,$-$20$^\circ$ }
\put( 060,200){\bf J2340$-$0053}
\put( 150,80) {\bf       CL = 1.2}
\put( 110,70) {\bf       B:0.018$^{\prime\prime}$$\times$0.005$^{\prime\prime}$,$-$18$^\circ$ }
\put( 240,200){\bf J2358$-$1020}
\put( 335,80){\bf       CL = 1.0}
\put( 290,70){\bf       B:0.021$^{\prime\prime}$$\times$0.005$^{\prime\prime}$,$-$19$^\circ$ }
\end{picture}
\label{VLBA_all}
\end{figure*}
\begin{figure*} 
\centering
\includegraphics[width=0.8\hsize,bb=18 18 594 774,clip=,angle=0]{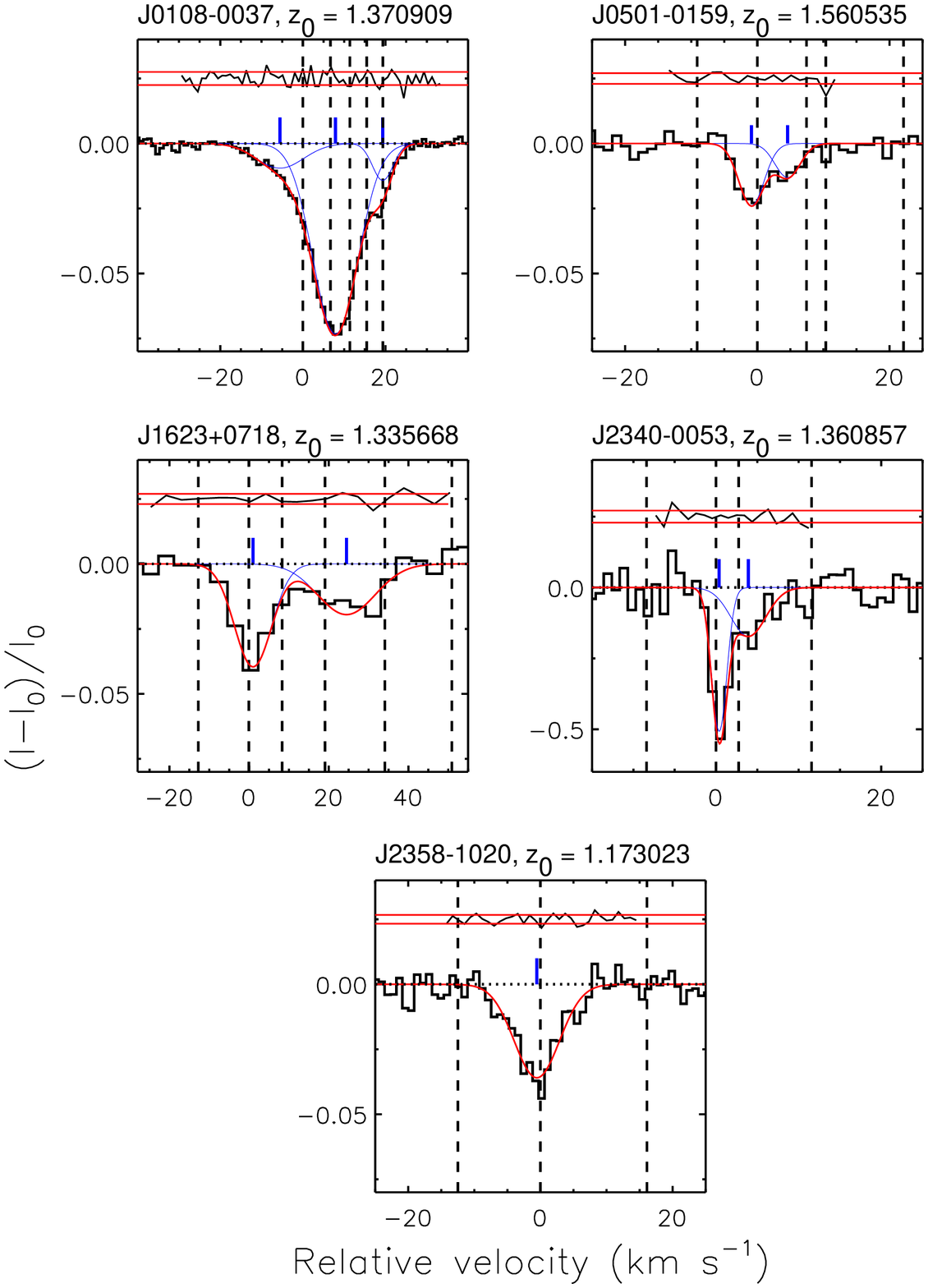}
\caption{21-cm absorption profiles of all the  \MgII\ absorbers in our sample. 
The redshift used to define the zero velocity scale is given in each panel as 
$\rm z_{0}$. The ticks mark the positions of the 21-cm absorption components. The long-dashed vertical lines 
are the positions of UV-optical absorber components (see Section~\ref{Voigt}). 
The normalized residuals (i.e. ([data]$-$[model]) / [error]) for each fit is shown  
in the top of each panel along with the 1$\sigma$ horizontal line.}
\label{fig_radio_all_v}
\end{figure*}
%
%
The five quasars being studied in this work seem to be 
compact in FIRST and GMRT 610-MHz images that both have 5 arcsec spatial resolution. 
VLBA  L-band 1422 MHz observations of 4 sources  (i.e. excluding J0501$-$0159) 
have been obtained as a part of a larger survey of radio sources 
with damped Lyman-$\alpha$ absorbers (DLAs) and \MgII\ absorption systems  along their line of sight 
to understand the relationship between radio structure and 
the detectability of 21-cm absorption \citep[see][]{Srianand12,Gupta12}. 
Details of observational setups and data reduction can be found in these papers.  
The final images of these 4 sources are shown in Fig.~\ref{VLBA_all}.

In Fig.~\ref{VLBA_all} we see that J0108$-$0037 is clearly resolved into several components 
at milliarcsec scales. We note that 73\% of the L-band flux density detected in the FIRST image is 
recovered in our VLBA image with the compact unresolved component having 53\% of the flux density. 
In the 8 GHz VLBA image the strong component seen in our L-band VLBA image gets
resolved into two distinct components and  only one of them may be associated to the optical continuum 
emitting region. This  makes our task  difficult while trying 
to assign 21-cm absorbers to their UV-optical absorbing counterparts as 
there is a high probability that additional contribution to the 21-cm 
absorption may come from  gas that is not located along the optical line of sight. 

From Fig.~\ref{VLBA_all}, we see that the source J1623$+$0718 is unresolved even at milliarcsec scales. 
However, the VLBA observation recovers only 42\% of the flux density detected in arcsec scale  FIRST image. 
Based on a Gaussian fit to our VLBA image we derive 
that the maximum angular extent of the compact component is 4.64 mas. 
This means the  size of the radio beam 
at the redshift of the absorber, $z$ $\sim$ 1.337, is less than 39 pc. 
It is likely that this radio source samples the same region of the absorbing gas as the optical source.
High frequency VLBA observations are not available in the literature for this source. In our
recent L-band GMRT observations we find that the flux density of this source at arcsec scales is 
similar to what is seen     in the FIRST image. This suggests that the $\sim$58 \% missing flux 
in our L-band VLBA image may be due to a diffuse extended radio emitting component that might have 
got resolved out at milliarcsec scale. Therefore if the absorbing gas is extended beyond 39 pc 
then the absorption against this diffuse component may also contribute to our GMRT spectrum.

The source J2340$-$0053 is clearly unresolved  in our L-band VLBA
image. It is also unresolved in the high frequency VLBA images taken at 
2 and 8 GHz \citep[for example see][]{Kovalev07}. 
The radio flux density shows a peak around 1.4 to 2.3 MHz with a sharp 
decrease towards low frequency end.  All this is consistent with the 
background quasar being a GHz peaked compact self-absorbed radio source. 
Using the flux density measurement in the FIRST catalog we find 90\% of the 
L-band flux density in the  FIRST image is recovered in the unresolved VLBA component. 
Gaussian  fitting of the L-band VLBA image gives the maximum angular extent 
of the object to be 1.71 mas. Using the redshift 
of the 21-cm absorber at $z\sim 1.36$ we estimate the maximum physical 
size of the quasar radio beam at the position of the absorbing gas to 
be $\le$15 pc. Thus it is most likely that optical and radio beams sample the same volume
of absorbing gas.  Therefore we expect the systematics 
related to the structure of the background radio source to   
be minimum in this case.

The source J2358$-$1020, observed with a resolution 
of $\sim$20 mas (Fig. \ref{VLBA_all}), is unresolved in our
L-band VLBA image. This source remains unresolved even in higher frequency 
VLBA observations taken at 2 and 8 GHz \citep{Fomalont00,Fey00}.
Similarly to the case of J2340$-$0053, this source is a GHz peaked 
radio source with a clear turnaround at the low frequency end. 
Moreover, in our VLBA observation we recover 74\% of the 
flux detected in  FIRST observation.  A single Gaussian component fit
represents its 1422~MHz VLBA image well. The largest angular 
size of the source is constrained to be $\le$3.71 mas which 
is equal to $\le$31 pc at the redshift of the absorbing system.
All this suggests that the optical and radio sight lines 
probe the same volume of the absorbing gas.  Therefore we 
expect the systematics related to the radio structure of this background 
quasar also be minimum.

The background radio source J0501$-$0159, also known as PKS~0458$-$020, exhibits multiple
components at arcsecond and milliarcsecond scales. The radio structure of this
source is investigated in detail by \citet{Briggs89} to determine the
spatial extent of the \zabs\ = 2.04 absorber 21-cm detected by \citep{Wolfe85}.  
At 1.6 GHz, the source is resolved into two components that are marginally
separated at 1 arcsec resolution \citep[see Fig.1 of][]{Briggs89}. 
In the 10 mas resolution map at 608 MHz, the frequency that is also close 
to the redshifted 21-cm frequency of the \zabs\ = 1.56 absorber, the compact
`core' at arcsecond scales is further resolved into a jet and a 
diffuse component \citep[see Fig.2 of][]{Briggs89}. The radio emission
in these different components is strong enough to contribute to the detected
21-cm absorption. Therefore, if the absorbing gas 
extends over several milliarcsecs, the possibility of velocity offsets between
radio and optical absorption lines due to the extended radio structure
cannot be ruled out in this case.     

\section{Gaussian fits to 21-cm absorption lines}
\begin{table*}
\begin{center}
\caption{Results of multi-component Gaussian fits to the 21-cm absorption line}
\begin{tabular}{lclcll}
\hline
\hline
Quasar &Channel width& \multicolumn{1}{c}{$z_{21}$}& \multicolumn{1}{c}{$\delta v$} &\multicolumn{1}{c}{$\delta v_{RR}$} &\multicolumn{1}{c}{$\delta v_{LL}$} \\
(1)&(2)&\multicolumn{1}{c}{(3)}&(4)&\multicolumn{1}{c}{(5)}&\multicolumn{1}{c}{(6)}\\
\hline
\multirow{3}{*}{J0108$-$0037}      & \multirow{3}{*}{1} &1.3708647(115)& 1.45 &\multicolumn{1}{c}{....}&\multicolumn{1}{c}{....} \\
                  &   &1.3709710(11)& 0.14   &\multicolumn{1}{c}{....}&\multicolumn{1}{c}{....}\\
                  &   &1.3710614(19)& 0.24   &\multicolumn{1}{c}{....}&\multicolumn{1}{c}{....}\\ 
\\
\multirow{2}{*}{J0501$-$0159}      &1  &1.5605277(32)$^\dagger$  & 0.38 & $-0.12\pm0.16^{\star}$  & $+0.30\pm0.44^{\star\star}$    \\
                  &1  &1.5605745(60)  & 0.70 & $+0.46\pm0.67^{\star}$  & $-0.16\pm0.40^{\star\star}$    \\
\\
\multirow{2}{*}{J1623$+$0718}      &\multirow{2}{*}{4}  & 1.3356761(51)$^\dagger$  & 0.65  & $+0.14\pm1.42$   & $-0.47\pm0.78$   \\
                                   &   &1.3358591(98)       & 1.26 &  $-1.60\pm3.70$ & $+0.76\pm2.41$  \\
\\
\multirow{2}{*}{J2340$-$0053}      &\multirow{2}{*}{1}  &1.3608595(14)$^\dagger$   & 0.18  & $+0.17\pm0.20$    & $+0.00\pm0.19$  \\ 
                                   &  &1.3608874(106)  & 1.35 &  $+0.42\pm0.48$   & $+0.39\pm2.60$  \\
\\
\multirow{2}{*}{J2358$-$1020}      &\multirow{2}{*}{1}  &1.1730206(17)            & 0.23  & $+0.74\pm0.51$   & $-0.55\pm0.28$  \\  
                                                     &  &1.1730188(25)$^\dagger$  & 0.35  & $+0.61\pm0.48$   & $-0.30\pm0.28$  \\ 
\hline
\end{tabular}
\begin{flushleft}
Column 1: Source name. Column 2: Channel width in \kms. Column 3: Absorption redshift and its error.
Column 4: The error in the absorption redshift in \kms. Column 5: Measured velocity   
offset with respect to the $z_{21}$ given in Column 3 in \kms\ when only RR spectrum considered. Column 6: Measured velocity  
offset in \kms\ when only LL spectrum considered.\\
$^{\star}$ XX   polarization in GBT; $^{\star\star}$ YY   polarization in GBT \\
$^\dagger$ component and corresponding $z_{21}$ used for measuring
$\Delta x/x$ \\
\end{flushleft}
\label{21cmgfit}
\end{center}
\end{table*}
%
%
%
\subsection{J0108$-$0037}
\citet{Gupta07} have reported 21-cm absorption from the \zabs\ = 1.371
\MgII\ system towards J0108$-$0037. Despite a low rest equivalent
width of $\sim$0.3 \AA\ for the \MgII\ doublet, 
other absorption lines of weaker metal transitions are clearly
seen even in the SDSS spectrum. Note that \zabs\ is very close
to \zem with an apparent ejection velocity of only $\sim$180 \kms.  
The 21-cm absorption is well approximated by a single Gaussian component in the GMRT data 
with a resolution of 4 \kms channel$^{-1}$. Subsequently, 
we observed this system for 25 hours spread over three observing runs of similar 
durations to obtain a spectrum with $~$1 \kms\ channel width. 
The final 21-cm absorption spectrum is the inverse-variance 
weighted mean of these 3 spectra (see top panel in 
Fig.~\ref{fig_radio_all_v}). It can be seen from the figure that the
absorption profile is smooth and that more than one Gaussian component is required to fit the profile. The smooth broad wing 
could indicate either a shallow component with a large velocity width or a blend 
of several weak narrow components  as suggested by the fit to metal absorption 
lines (see section \ref{0108_opt}). 
 We find that at least three components are needed to obtain a good
fit with a reduced $\chi^2$ of 1.1. The result 
of this fit is over-plotted on top of the observed spectrum in Fig.~\ref{fig_radio_all_v}. 
The redshift of the strongest 21-cm component is 1.3709710(11) (see Table~\ref{21cmgfit}). 
The reduced $\chi^2$ is 1.03 for a 4 component fit. This suggests that four components 
are adequate to represent the 21-cm profile. The redshift of the strongest 21-cm 
component in this case is 1.3709694(53). 
\subsection{J0501$-$0159}
The sight line towards this quasar is interesting as it covers 
two 21-cm absorbers at $z$ = 2.04 and 1.56 \citep[][]{Wolfe85,Kanekar09}. 
\citet{Kanekar10} fitted the 21-cm absorption with a single component at
$z_{21}$ = 1.5605300(25) in a spectrum smoothed to a resolution of 
1.3 \kms channel$^{-1}$. Using the pipeline based on the NRAO's GBTIDL package
we re-reduced their archived GBT data. Details on the GBT data reduction can
be found in \citet{Srianand12}. The original data was obtained with  
two sets of spectral resolutions (i.e. 0.33 and 0.66 \kms\ channel$^{-1}$).
However, we rebinned the individual spectra to 1.3 \kms\ channel$^{-1}$ 
resolution before combining them.  This is done to match the resolution 
to our GMRT spectra of other sources.
The absorption profile of this 21-cm absorber (shown in the top panel 
of Fig. \ref{fig_radio_all_v}) shows the existence of two absorbing 
components. Redshifts obtained from the two components fit are summarized in 
Table~\ref{21cmgfit}. It can be seen that the spectra obtained in two 
polarization channels give consistent redshifts for the two components 
within measurement uncertainties (i.e 380 and 700 m~s$^{-1}$ for 
the strong and the weak component respectively).
%
%
\begin{figure} 
\centering
\includegraphics[width=0.8\hsize,bb=18 18 594 774,clip=,angle=90]{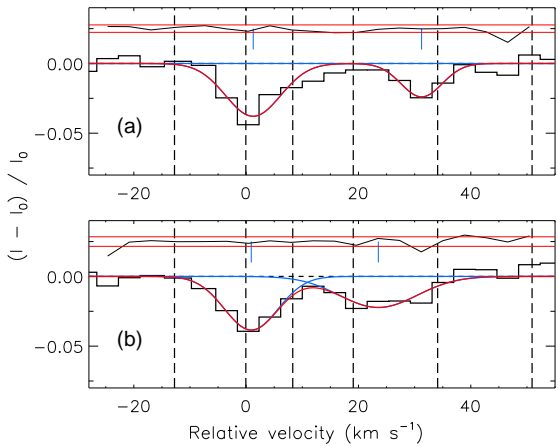}
\caption{New and old combined 21-cm spectra of J1623$+$0718 respectively 
in panel (a) and (b) with the best fitted double Gaussian in each case shown 
as continuous curve. The thick  histogram in the bottom of two panels 
show the residual of the fits that are shifted by a constant offset for clarity.  
The $v = 0$ is at $z$ = 1.335668. Although the redshift of the first component in both new 
and old spectra matches very well  the redshift of the second component (at $v \sim 35$ \kms in the top panel) 
differs  by 7.7$\pm$2.0 \kms\ between the two spectra.
The normalized residuals (i.e. ([data]$-$[model]) / [error]) for each fit is shown  
in the top of each panel along with the 1$\sigma$ horizontal line.}
\label{fig_oldnew_1623}
\end{figure}
%
\subsection{J1623$+$0718}
The 21-cm absorption at $z~\sim$ 1.336 towards J1623$+$0718 was first reported
by \citet{Gupta09} with two components.
Using their spectrum we measure $z_{21}$ = 1.3356755(53) and 
1.3358518(108). As the background source as well as the absorption 
lines are weak we reobserved this system for 16.8 hours spread over two 
full synthesis with a channel width of 2 \kms.
Unfortunately the SNR in the final spectrum is not high enough to provide accurate redshift 
measurements. Therefore, we rebinned our new spectrum to a resolution of
4 \kms\ channel$^{-1}$. The redshifts of the two components
in the new combined spectrum are  $z_{21}$ = 1.3356782(85) and
1.3359118(116). While the redshift of the blue component
is consistent with the measurement based on the spectrum of \citet{Gupta09}
the redshift of the second component is off by
7.7$\pm$2.0 \kms\ (see Fig. \ref{fig_oldnew_1623}). 
We attribute this to the low optical depth in this
component or to the presence of a low level RFI affecting the shallow feature. 
Because of this reason we do not use this component to constrain $\Delta x/x$.
In order to increase the SNR further, we combined our new spectra with the 
spectra obtained by \citet{Gupta09}. The Gaussian fit to the combined spectrum
is shown in Fig. \ref{fig_radio_all_v}. The fit results are summarized in 
Table~\ref{21cmgfit}. The measured $z_{21}$ = 1.3356761(51)  for the main 21-cm component 
agrees well with the measurements based on spectra obtained in 
individual polarization channels. The redshift uncertainty in this case corresponds to 
a velocity 
of 650 \ms.
\subsection{J2340$-$0053}
The 21-cm absorption at $z$ = 1.3606 along the line of sight towards this quasar was discovered 
by \citet{Gupta09}. We have acquired additional 3x8 hours of GMRT data at 1 \kms\ 
channel$^{-1}$ resolution. However  these  data  are found to be unusable due to  RFI. 
So we use here only the 
the GMRT 21-cm absorption spectrum with, $\delta v \sim$ 1 \kms\ channel$^{-1}$, obtained by 
\citet{Gupta09} and shown in Fig. \ref{fig_radio_all_v}. 
The spectrum clearly shows two absorbing components, the bluer  being quite strong. 
The continuous line in Fig~\ref{fig_radio_all_v} shows the double Gaussian fit to the 21-cm absorption profile. 
The two components are also shown (see also Table~\ref{21cmgfit}).  
The redshifts of the two components are $z$=1.3608595(14) and $z$=1.3608874(106). The redshift errors 
for the Gaussian fits correspond to uncertainties in the velocity scale of 180 \ms\ and 1350 \ms\  
respectively. The above redshifts are found to be consistent with those obtained by 
fitting  RR and LL spectra separately (see Table~\ref{21cmgfit}). 
\subsection{J2358$-$1020}
\begin{figure*} 
\centering
\includegraphics[width=0.85\hsize,bb=18 18 594 740,clip=,angle=90]{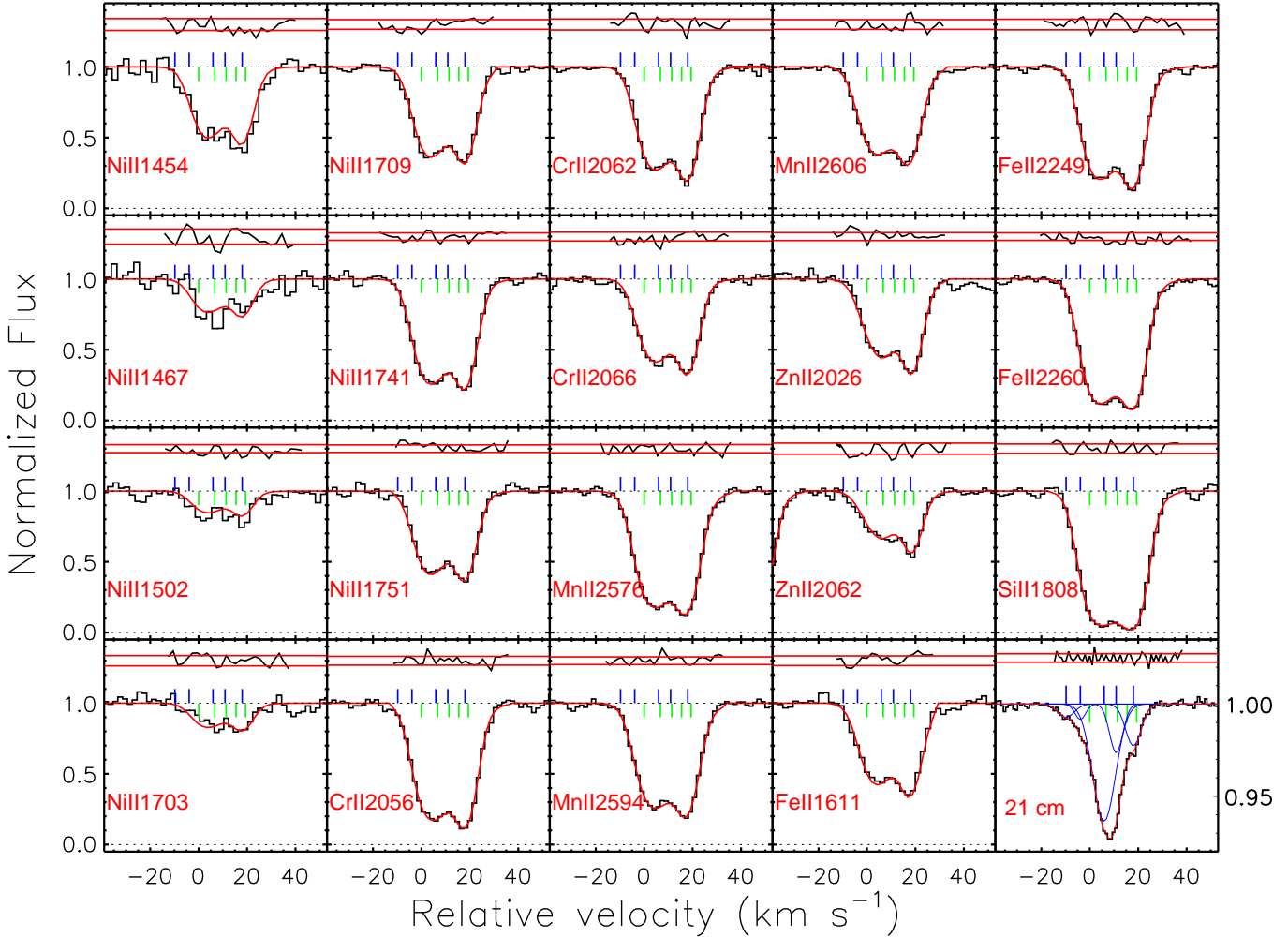}
\caption{Voigt profile fits to the absorption profiles of the \zabs\ $\sim$ 1.37 system  
towards J0108$-$0037. 
The histogram plot in each panel shows the observed absorption profile of a given transition 
and the continuous curve is the best Voigt profile fit (or Gaussian fit in the 
case of 21-cm profiles). Last panel shows the 21-cm absorption profile along with the fitted model and 
its individual components.  The normalized residuals (i.e. ([data]$-$[model]) / [flux error]) for each fit is shown  
in the top of each panel along with the 1$\sigma$ horizontal line.} 
Upper vertical tick marks indicate the position of 21-cm absorber components and lower vertical 
tick marks indicate different optical-UV velocity component. 
\label{fig0108_ionII}
\end{figure*}

The 21-cm absorption along the line of sight towards this quasar was 
first discovered by \citet{Gupta07}. The 21-cm absorption is 
well approximated by a single Gaussian component at a spectral
resolution of 2 \kms channel$^{-1}$ \citep{Gupta09}. 
Two high resolution spectra of this object with 
$\delta v \sim$ 1 \kms\ per channel were acquired during subsequent observations (2$\times8$ hours). 
The shape of the 21-cm absorption feature even at this higher resolution is consistent with a 
single component (see Fig.~\ref{fig_radio_all_v}). This is the simplest profile in our sample.
It is best fitted with a single  component at $z_{21}$ = 1.1730206(17).
The typical error in the redshift measurement is $\sim$230 m~s$^{-1}$.
As there are 4 individual spectra we also performed a fit using errors that are 
the rms of the fluxes measured in the different spectra.
The fit obtained using these errors gives  $z_{21}$ = 1.1730212(20) which is consistent
with the above quoted measurement. In Table~\ref{21cmgfit} we also provide
results of independent fits to RR and LL spectra. The $z_{21}$ measurement
based on the RR spectrum is higher than the one obtained from the LL spectrum with a 
relative off-set of 1.3$\pm$0.6 \kms. We find this is mainly due to one of the RR spectra being 
affected by low level RFI. While this spectrum does not influence the
weighted mean I-spectra, the combined RR spectrum is appreciably affected by this.
Combining all the spectra but this affected spectrum yields
$z_{21}$ = 1.1730188(25) (i.e with a redshift error corresponding to 344 m~s$^{-1}$). 
This is the $z_{21}$  we use to derive \dxx\ for this system. 
\section{Voigt profile fitting of UV lines}
\label{Voigt}
In this section we describe the Voigt profile fitting of the UV absorption
lines and discuss the individual systems in detail.

\begin{figure*} 
\centering
\includegraphics[width=0.85\hsize,bb=18 18 594 755,clip=,angle=90]{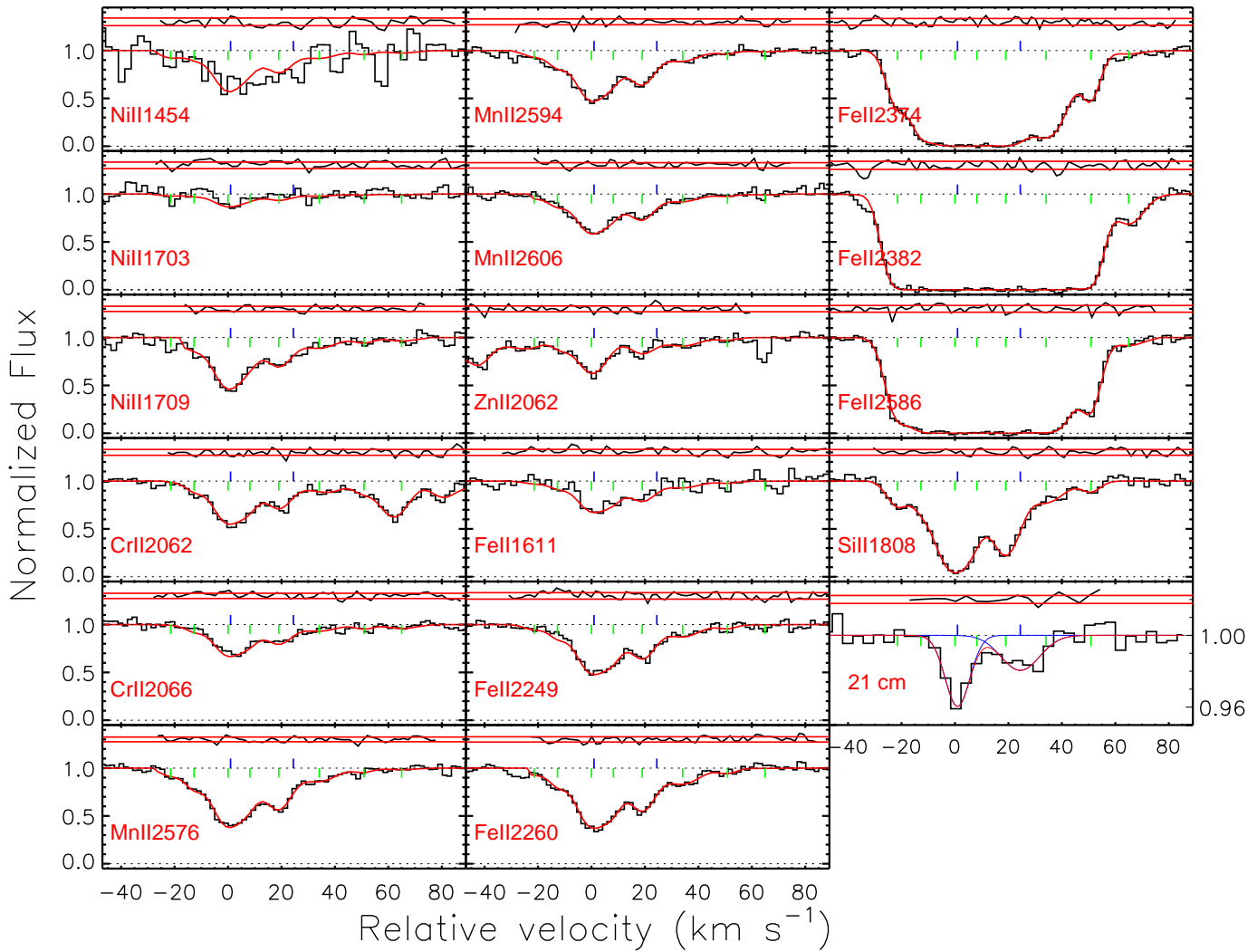}
\caption{Same as Fig. \ref{fig0108_ionII} for \zabs\ $\sim$ 1.33 \MgII\ system 
towards J1623$+$0718. 
}
\label{fig1623_ionII}
\end{figure*}

\subsection{System at \zabs\ $\sim$ 1.37 towards J0108$-$0037}\label{0108_opt}
J0108$-$0037 is one of the brightest quasar in our sample with SDSS r-band 
magnitude of 17.5. The SNR in the continuum close to the absorption lines used for 
 measuring redshifts is usually larger than 30  for this system. 
The absorption profiles of all singly ionized transitions 
used in the Voigt profile fits are shown in Fig.~\ref{fig0108_ionII}. 
Interestingly  many transitions of \NiII\ are detected in this system. 
This could allow one to measure \daa\ by using only \NiII\ transitions once 
accurate values of rest frame wavelengths, oscillator strengths, and sensitivity coefficients are available. 
Apart from \NiII$\lambda$1467,1502,1703, other absorption profiles are very strong. 
Absorption profiles of \FeII$\lambda$1608,2344,2374,2382,2586,2600 in our final combined UVES spectra
are highly saturated and have not been used for redshift measurement. 
We have fitted this system with two, three, four, and five components to find the optimal fit. 
The reduced $\chi^2$ are respectively, 1.60, 1.42, 1.28, and 1.23. Increasing the number of 
components does not lead to any better fit. Therefore we consider 
the fit with five components as the best fit for this absorbing system. 
 From our best fit, there are two
strong UV absorption components (at $v$ = 6.4 and 19.5 \kms\ in 
Fig.~\ref{fig0108_ionII}) with approximately the same column 
density of metals. 
This means, unlike in other cases discussed 
here, a unique identification of the stongest UV component to be
associated with the strongest 21~cm component is highly
questionable. 
Even though both 21-cm and UV absorption lines span the same velocity range 
there is no one to one correspondence between the two. 
This could mean that the 21-cm optical depth does not scale with 
the column density of metal lines. To illustrate this we plot
in the bottom right panel of  Fig. \ref{fig0108_ionII} a five
component fit of the 21~cm absorption profile.
This could mean as well that additional contribution to 21-cm 
absorption comes from gas that is not probed by the optical sight line. 
As the morphology of the background source is complex, we can not
rule out that the differences in the absorption profiles are due to
the fact that the optical and radio sight lines probe 
different volumes of the absorbing gas. Therefore,
because of the degeneracy introduced by this peculiar profile
and the complex morphology of the radio emission we do not use this 
system for \dxx\ measurements.
%
%
%
\begin{figure*} 
\centering
\includegraphics[totalheight=.75\textheight,width=0.70\textheight,bb=18 18 594 755,clip=,angle=90]{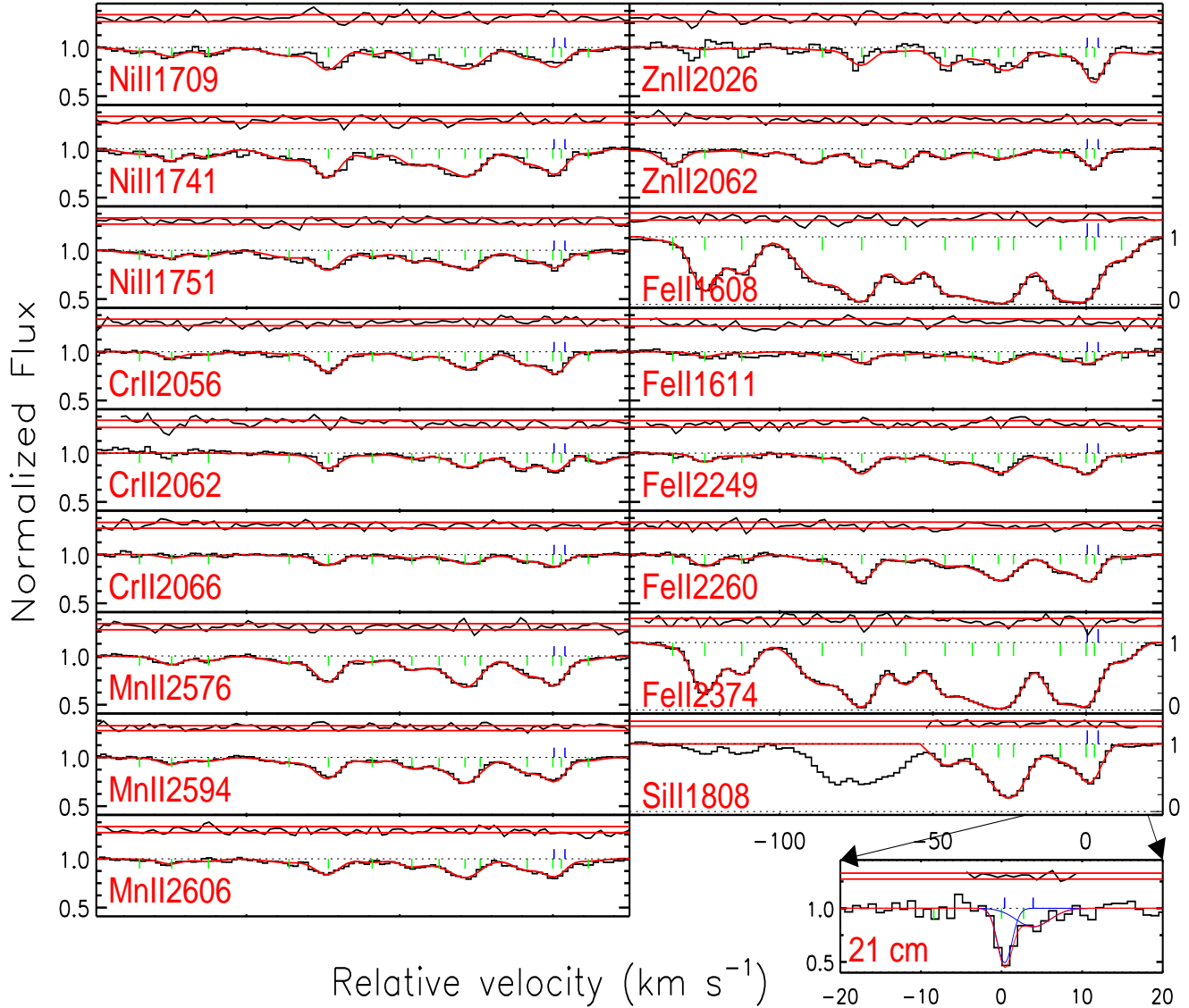}
\caption{ Same as Fig. \ref{fig0108_ionII} for the \zabs$\sim$1.36 \MgII\ absorber towards J2340$-$0053 
and for the VLT/UVES. In the bottom right corner panel we zoom over 40 \kms\ to show the 21-cm absorber with 
the two components fit overplotted.
}
\label{fig2340_ionII}
\end{figure*}
%
%
\subsection{System at $z$ $\sim$ 1.33 towards J1623$+$0718}
J1623$+$0718 with SDSS r-band magnitude of $\sim$17.5 is another bright 
quasar in our sample. The velocity plots of some of the species detected from the \zabs$\sim$1.33 
system are shown in Fig.~\ref{fig1623_ionII}. The SNR in the continuum close to 
absorption profile of \NiII$\lambda$1454 is $\sim$10, 
and it is  higher than 30 close to \FeII$\lambda$2586. The absorption 
profiles of weak metal transitions like \SiII$\lambda$1808  are 
spread over 80 \kms. The 21-cm absorption including the broad component
is spread over the same velocity range. From visual inspection it is 
clear that the strongest metal absorption component coincides 
well with the main narrow 21-cm component. 

The  optical absorption profiles suggest the presence of  additional weak components in
the wings (at $v$$\sim$53 and $\sim$$-20$ \kms). We constrain the component structure in the wings 
from the strong \FeII\  lines for which the wings are apparent.
The best fitted Voigt profile shown in Fig. \ref{fig1623_ionII} has a reduced $\chi^2$ of 
1.04. 
The measured redshift of the  strongest metal component in the 21-cm velocity range 
is \zabs\ = 1.3356684(19) with the typical redshift
error of 240 \ms\ (see Table~\ref{tab_sys_all}). 
\begin{figure*} 
\centering
\includegraphics[totalheight=.75\textheight,width=0.70\textheight,bb=18 18 594 774,clip=,angle=90]{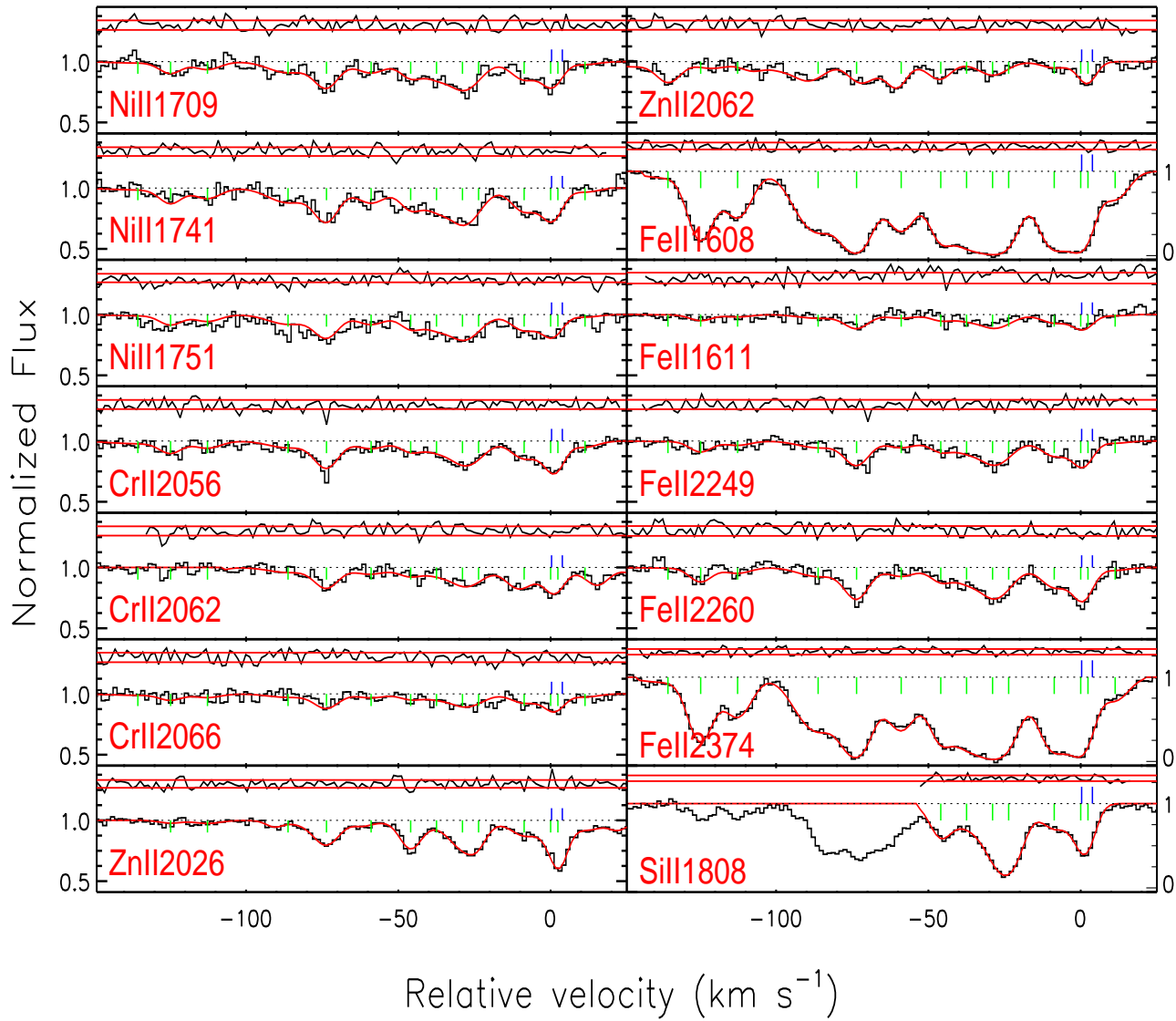}
\caption{Same as Fig. \ref{fig0108_ionII} for the \zabs$\sim$1.36 \MgII\ absorber towards J2340$-$0053 
and for Keck/HIRES. 
}
\label{fig2340_ionII_keck}
\end{figure*}

As can be seen, the main component in the metal absorption is broad and may 
contain additional hidden narrow components. Therefore, we repeated the fits with additional components 
injected around the main component. 
The reduced $\chi^2$ does not change with the addition of these new components.
There is however a minor change in the absorption redshifts, albeit with increased errors,
for the strongest component we find $z$ = 1.3356675(47) with an error of $\sim$600 \ms.
This is the UV absorption redshift and associated uncertainty we consider for \dxx\ measurement.
%
%
\begin{figure*} 
\centering
\includegraphics[width=0.58\hsize,bb=90 18 522 774,clip=,angle=90]{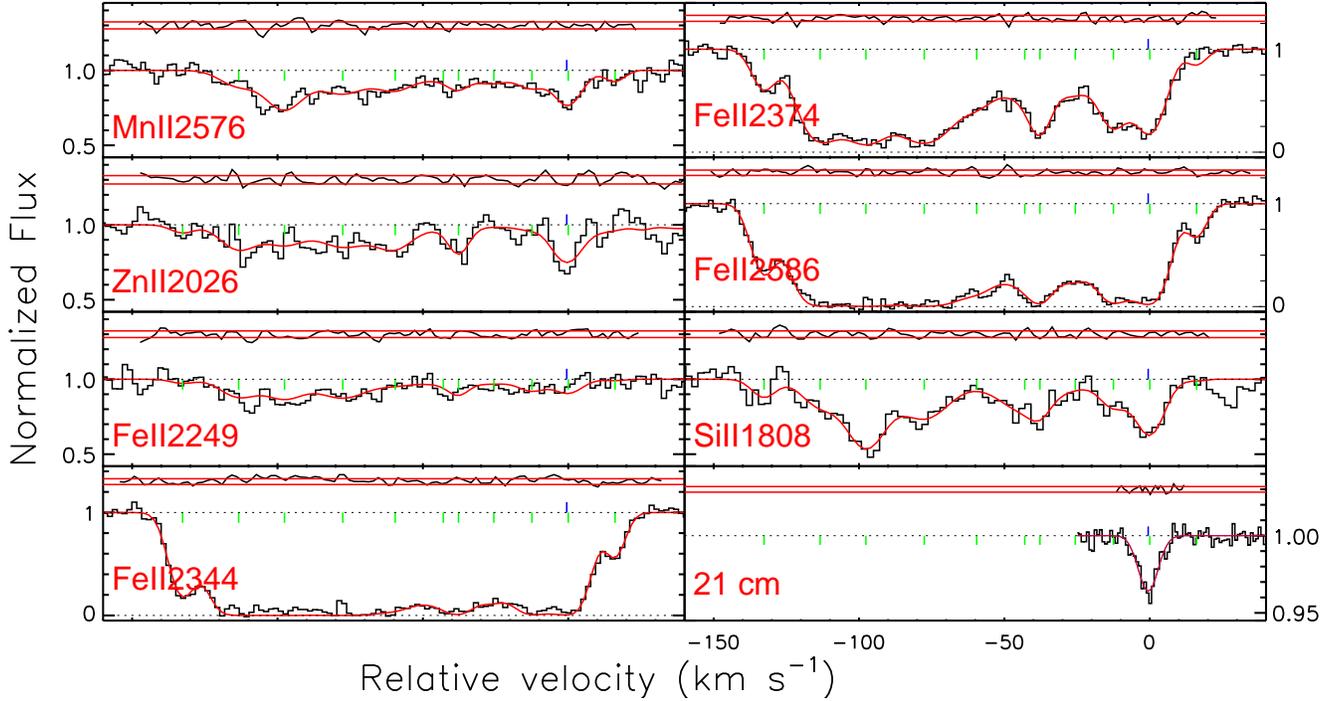}
\caption{ Same as Fig. \ref{fig0108_ionII} for the \zabs$\sim$1.17 \MgII\ absorber towards J2358$-$1020. 
}
\label{fig2358_ionII}
\end{figure*}
\subsection{System at \zabs$\sim$1.36 towards J2340$-$0053}
In addition to our VLT/UVES spectrum we have also analysed the
Keck/HIRES spectrum of this object. The velocity plot of different
ions detected in this system is shown in Fig.~\ref{fig2340_ionII} and Fig.~\ref{fig2340_ionII_keck}.
The SNR in the UVES continuum is $\sim$40 close to \NiII$\lambda$1709 and 
can be larger than 60 close to \MnII\ absorptions.  
The absorption lines of singly ionized species are spread over $\sim$150 \kms. Unlike 
in the case of J1623$+$0718 and J0108$-$0037 the 21-cm absorption is very much narrower
than the UV absorption lines. However, as pointed out by \citet{Gupta09},
the metal component associated with the 21-cm component is well
detached. In particular the component is well defined by \SiII$\lambda$1808
and \ZnII\ lines. Note, we have not fitted the absorption profile 
of \SiII\ for $v < -55$ \kms\ as this region is contaminated by 
absorption from another intervening system. The fit has a reduced $\chi^2$ of 1.3. 
The overall profile is fitted with 14 Voigt profile
components and the absorption coinciding with the 21-cm absorption
requires two narrow components.

The redshifts of the two UV-optical components 
in our UVES spectrum that are closer to 21-cm absorption are 1.3608565(16) and
1.3608781(19). The former happens to be the stronger absorption 
component to be associated to the stronger 21-cm absorption component. 
The redshift errors due to Voigt profile fitting correspond to 
200 and 240 \ms\ for these components. The redshifts 
measured from the HIRES spectrum, 1.3608572(36) and 1.3608759(40), are consistent
with those derived from the UVES spectrum. 
The fitting errors in the redshifts for the HIRES data are 457 and 508 \ms\ respectively.
%
%
\subsection{System at \zabs\ $\sim$ 1.17 towards J2358$-$1020}
J2358$-$1020 with an r-band SDSS magnitude of 18.7 is  one of the faintest quasars in our 
sample. 
The SNR in the UVES spectrum is at best $\lesssim$25. 
As it can be seen in Fig.~\ref{fig2358_ionII} the absorption profiles 
are spanning more than 150 \kms.
Absorption profiles of \CrII$\lambda$2056,2062,2066, \ZnII$\lambda$2062, 
\MnII$\lambda$2594,2606 are not used in the fit as they are weak or located in regions of poor SNR.
As there are weak components at v$\sim$$-$140 and 15 \kms, we included two core-saturated profiles,
\FeII$\lambda$2344 and \FeII$\lambda$2586, to be able to fit the overall profile.
Similar to the case of J2340$-$0053 the metal component coinciding with the 21-cm absorption
seems to be well detached from other components  and is well fitted with a 
single Voigt profile component. This is apparent for \SiII$\lambda$1808, 
\ZnII$\lambda$2026, and  \MnII$\lambda$2576. The overall fit has 11 Voigt profile 
components with a reduced $\chi^2$ of 1.2. 

The redshift of the stronger absorption component that we associate 
 with the 21-cm absorption is 1.1730227(29). The redshift error from 
line fitting is $\sim$400~\ms. 
%
\subsection{System at \zabs\ $\sim$ 1.56 towards J0501$-$0159}
\begin{figure*} 
\centering
\includegraphics[totalheight=.75\textheight,width=0.65\textheight,,bb=18 18 594 764,clip=,angle=90]{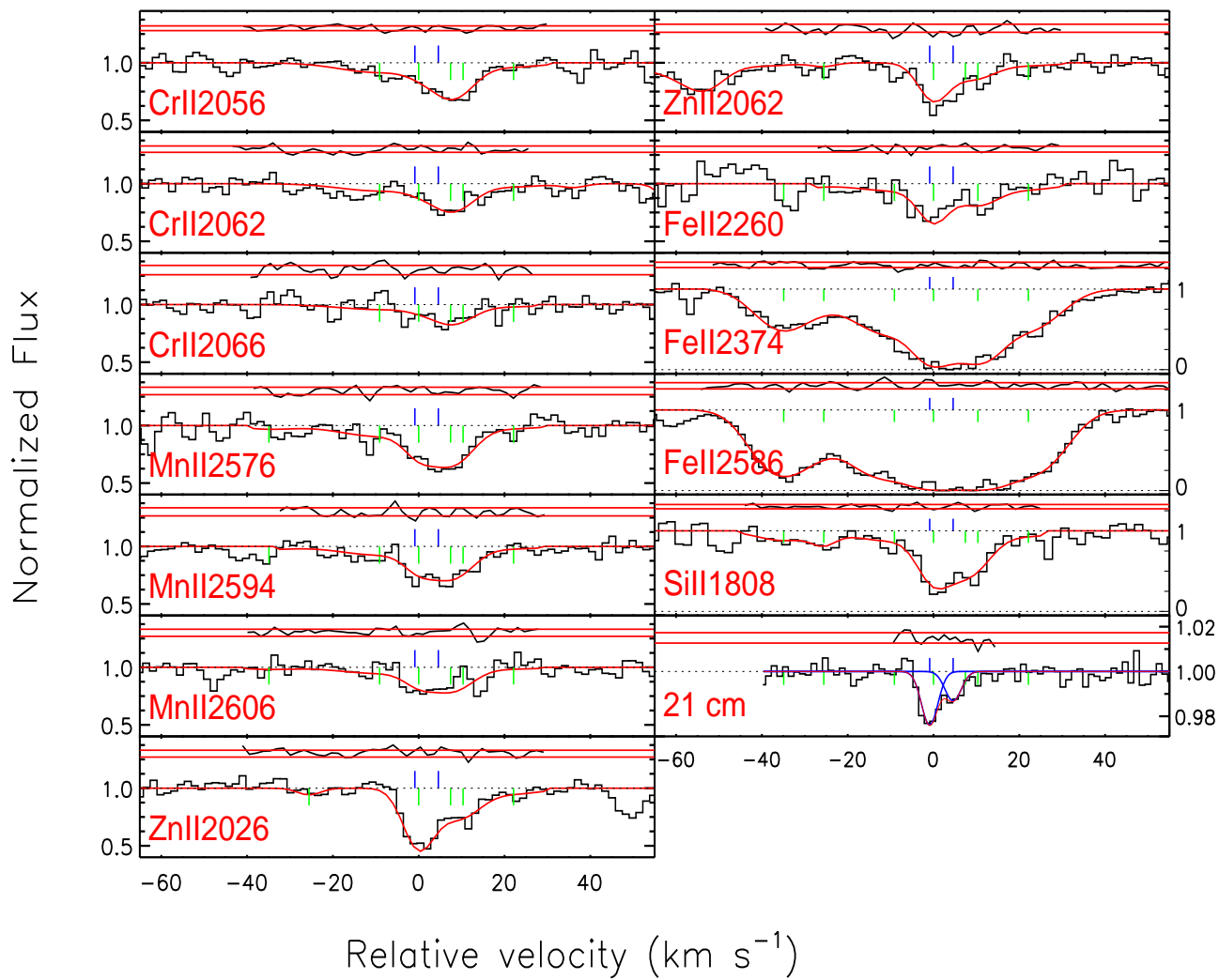}
\includegraphics[totalheight=.71\textheight,width=0.29\textheight,bb=108 18 504 774,clip=,angle=90]{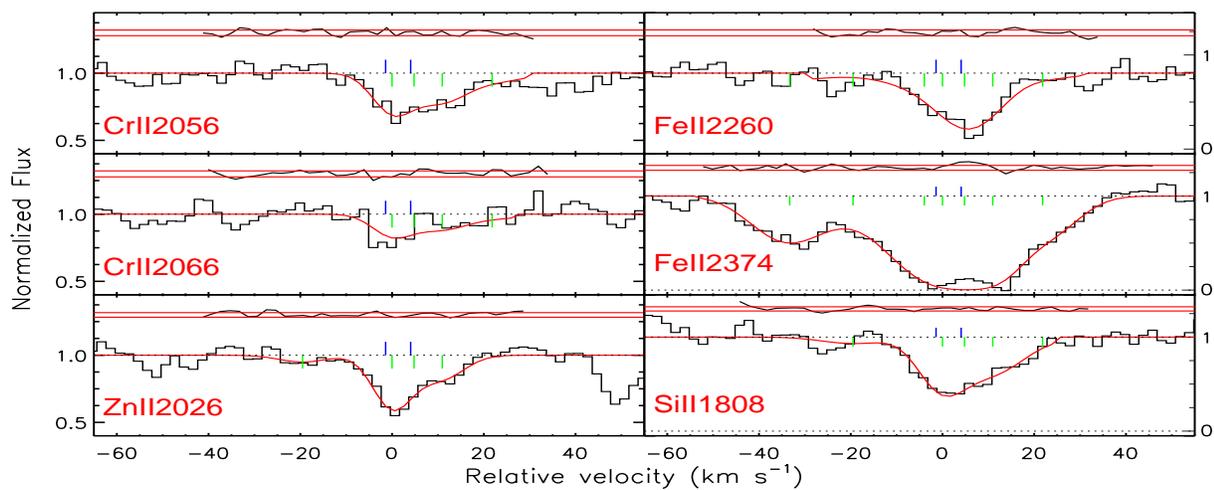}
\caption{Same as Fig. \ref{fig0108_ionII} for the \zabs$\sim$1.56 \MgII\ absorber towards J0501$-$0159
from VLT/UVES spectrum (top) and HIRES/Keck spectrum (bottom).}
\label{figpks_ionII}
\end{figure*}

J0501$-$0159 is a faint quasar with an r-band magnitude of 19.33. Similar 
to the case of  J2340$-$0053 we  have spectra obtained with VLT/UVES as well as Keck/HIRES. 
Fig.~\ref{figpks_ionII} presents the velocity plot of different absorption profiles detected in this system. 
The top and bottom panels show the VLT/UVES and Keck/HIRES spectra respectively, along with their best fit 
profiles. Although the final combined UVES/VLT spectrum is made of 10 exposures each 
of more than 3300~s of exposure time, the typical  SNR is only $\lesssim$20.
Apart from those \FeII\ lines that are shown in  Fig. \ref{figpks_ionII} 
other \FeII\ lines are highly saturated or have poor SNR. Associated \NiII\ absorption lines are very weak and are 
not used for the Voigt profile fitting. Some of them are located in the Lyman-$\alpha$ forest, others are either contaminated 
or have poor SNR and therefore have not been included in the fit.  
Our best fit is shown in Fig. \ref{figpks_ionII} and has a reduced $\chi^2$ of 1.0. 

It can be seen in Fig.~\ref{figpks_ionII} that the strongest 21-cm component
coincides with the  strongest metal component seen in the 
absorption profiles of  undepleted species like \ZnII\ and \SiII.
The redshifts of the UV-optical component in UVES and HIRES data are, respectively,
1.5605354(43) and 1.5605398(276) with errors of 503 \ms\ and  3234 \ms. 
The two measurements agree within 1.0$\sigma$.
%
\section{Constraining \dxx}
\begin{figure} 
\centering
\includegraphics[width=1.1\hsize,bb=18 17 594 773,clip=,angle=0]{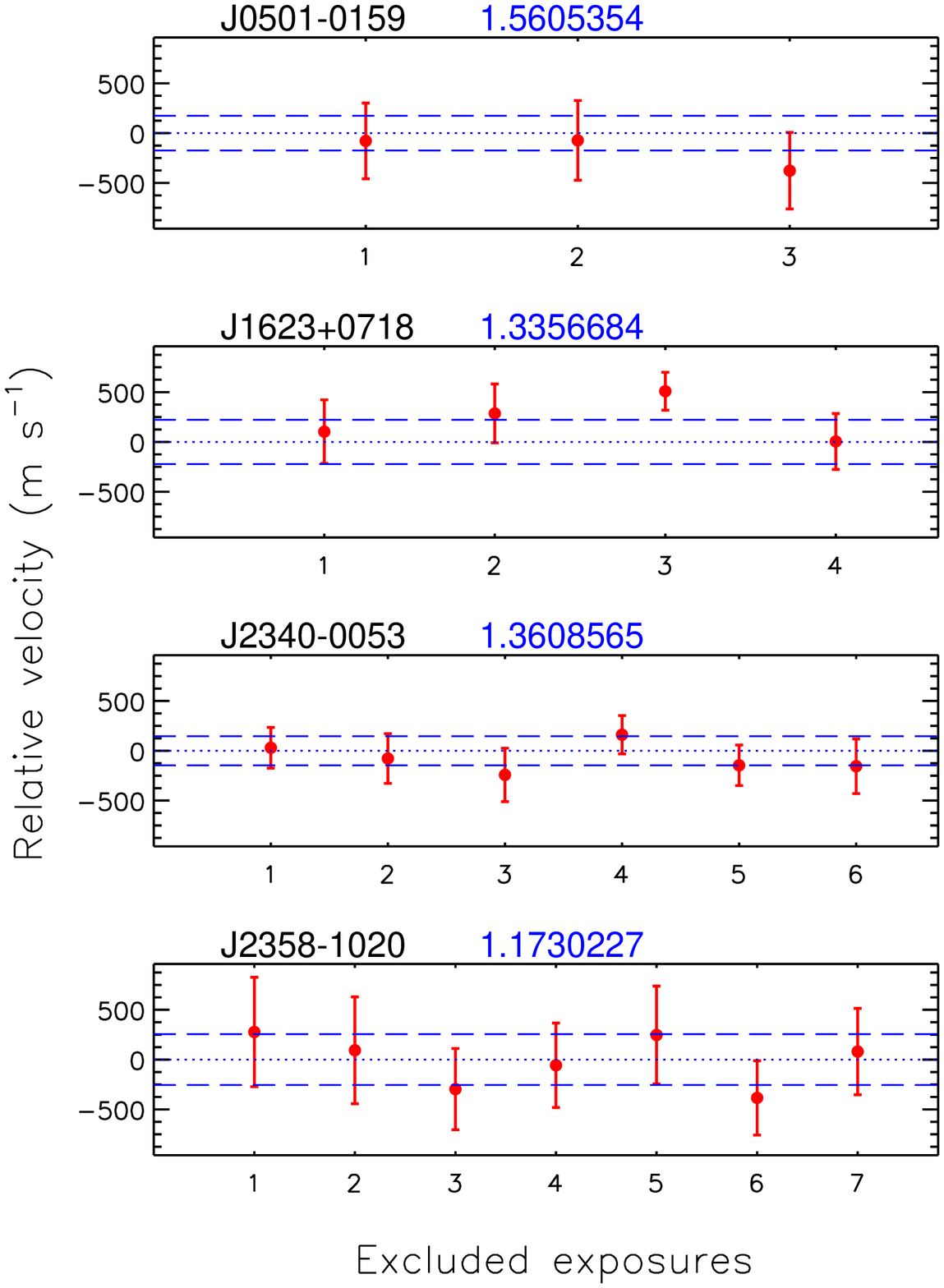}  
\caption{
Results of an experiment consisting of measuring the shifts in the observed redshift  
as a consequence of retrieving one exposure when combining the different exposures
to obtain the final quasar spectrum. We only show those components that are 
used for measuring \dxx.  
Abscissa in each panel indicates the missing exposure(s). The zero of the y-axis is taken at the redshift
(indicated at the top of each panel)
measured in the spectrum obtained by combining all exposures (dotted line). Short-dashed lines 
shows the 1$\sigma$ error on this redshift.
}
\label{fig_miss_all}
\end{figure}
\begin{table*}
\caption{Results of repeated Voigt profile fitting analysis.
}
\resizebox {\textwidth }{!}{ %
\begin{tabular}{lcccccccccc}
\hline
\hline
Quasar & $z_{\rm UV}\dagger$ & $\delta v$   &  $\delta v_1^\dagger$  &   id &  $\delta v_2^\dagger$  &  id  & $\delta v_3^\dagger$ &  id &  $\delta v_4^\dagger$  &  $\delta v_5^\dagger$\\
     &  & (\kms) & (\kms) &&(\kms) &&(\kms) &&(\kms)&(\kms)\\
(1) & (2) & (3)& (4) & (5) & (6) & (7) & (8) & (9) & (10) & (11)\\
\hline

J1623+0718 & 1.3356684(19) & 0.24    & $-$0.02$\pm$0.23 & d2,e1,e5 & $+$0.34$\pm$0.31 & d5 & $+$0.36$\pm$0.26 &d2,d3,d7,d8,e5,c2 &  ---- & $+$0.23$\pm$0.22  \\
\\
%
J2340-0053 & 1.3608565(16) & 0.20    & $+$0.26$\pm$0.18 & d2,f1 & $-$0.22$\pm$0.19 & d8    &$-$0.48$\pm$0.33 & e7,a,d3,c1 & $+$0.09$\pm$0.46& $-$0.07$\pm$0.15 \\
\\
J2358-1020 & 1.1730227(29) & 0.40    & $+$0.26$\pm$0.47 & f1,d4 & $-$0.01$\pm$0.37 & d9 &$-$0.54$\pm$0.44 & d5,d6 &  ----  & $-$0.02$\pm$0.23\\
\\
J0501$-$0159& 1.5605354(43)  & 0.50  & $+$0.35$\pm$0.27 & b2,b3,d4,c3  &  $+$0.26$\pm$0.30 & d9 & $+$0.15$\pm$0.67 &b3,a,d3,d8,c2 &  $-$0.52$\pm$3.23& $+$0.07$\pm$0.17 \\
%
\hline
\end{tabular}}
\begin{flushleft}
Column 1: Source name. Column 2:  The absorption redshifts and associated
error (in brackets) measured using our VLT/UVES spectrum. Column 3: error
in the redshift measurement given in Column 2 in \kms. Column 4:
measured velocity offset when excluding weak transitions (as defined
in Table.~\ref{tabatomic}) listed in Column 5. Column 6: measured velocity
offset when a saturated line (with ids given in column 7) is included
in the fit. Column 8: Velocity offset measured after excluding the
absorption lines (whose ids are given in column 9) far away from the
ThAr lamp lines used for wavelength calibration. Column 10: Measured
velocity offset for the redshift measured using Keck/HIRES spectrum. 
Column 11: Mean and standard deviation of measured redshifts after removing 
one exposure from combined spectra (see Fig. \ref{fig_miss_all}).\\
$^\dagger$ all the velocities and associated errors are calculated with respect to the main redshift 
given in the second column. 
\end{flushleft}
\label{tab_sys_all}
\end{table*}

In this section we present \dxx\ measurements for individual
systems and discuss the associated errors in detail. We measure \dxx\ using,
\begin{equation}
{\Delta x \over x} = \bigg{(}{{z_{\rm UV} - z_{21}}\over  1+z_{21}}\bigg{)},
\end{equation}
where $z_{\rm UV}$ is the redshift 
of the UV-optical component as given in column 2 of Table~\ref{tab_sys_all} and $z_{21}$ is the
redshift of the the 21-cm component as given in column 3 of Table~\ref{21cmgfit}. 
Usually the metal absorption has a larger velocity spread compared to that of 21-cm 
absorption. So to associate the UV-optical absorption (i.e. $z_{\rm UV}$) to the 21-cm component 
we use the stronger absorption component in the velocity range of 21-cm absorption. In all 
cases discussed here the UV-optical absorption clump associated to the 21-cm 
absorption is easily identifiable.  
The statistical error due to the fits, $\sigma(\frac{\Delta x}{x})$, is calculated  as: 
\begin{equation}
\sigma(\frac{\Delta x}{x}) = \frac{1}{1+z_{21}}\sqrt{\left(\left(\frac{1+z_{UV}}{1+z_{21}}\right)^2\times\sigma_{z_{21}}^2 + 
\sigma_{z_{UV}}^2 \right) },
\end{equation}
in which $\sigma_{z_{\rm UV}}$ and $\sigma_{z_{21}}$ are the errors on $z_{\rm UV}$ and $z_{21}$ respectively. 
In the case of $\sigma_{z_{\rm 21}}$ we consider the contributions of statistical  (Table. \ref{21cmgfit})
and systematic uncertainties (Table~\ref{tabshift_21}), the latter being smaller
($\sim$ 122 \ms) than the former ($\geq$ 300 \ms).
To estimate the systematic errors in the optical redshifts  we carry out 
a number of tests whose results are summarized in Table~\ref{tab_sys_all} and also in the tables 
of the Appendix (see Section \ref{MgIIsample_optical} for more detail). 
In columns (4), (6), and (8) of Table~\ref{tab_sys_all} we give velocity offsets measured with 
respect to \zabs\ given in column (2) using repeated Voigt profile fitting after, respectively, 
excluding weak lines, including saturated lines, and excluding the  absorption lines 
with no ThAr line within 50 \kms\ of that  optical component assigned to 21-cm component. 
The reference to the corresponding lines are given in the preceding column.
The results of these tests are sensitive to the intra-order wavelength calibration errors. To be on 
the conservative side we consider the maximum error found here as a measure of the 
systematic error introduced from intra-order shifts ($\sigma_{\rm e}$). 

In column (11) of Table~\ref{tab_sys_all} we give the mean velocity offset found by repeated Voigt 
profile fitting of lines after excluding one of the exposures (See also Fig.~\ref{fig_miss_all}). Tables in the Appendix 
also summarize the results of cross-correlation analysis between the individual exposures and the 
combined spectra. The results of the last two exercises  (exposure removal and cross-correlation)
are sensitive to any constant shift between different exposures. Therefore we consider the maximum of 
these two shifts as an estimate of the constant shift error in the wavelength calibration ($\sigma_{\rm c}$). 
The final systematic error for each sight line is taken as the  quadratic sum of the two systematic errors.
The approach taken here is conservative to allow for maximum uncertainty in the measurement of $z_{\rm UV}$. 
We summarize the values of $\sigma_{\rm e}$ and $\sigma_{\rm c}$ 
for individual systems in Table \ref{tab_dxx_all} where these errors are converted 
from velocity shifts to \dxx. 
In column (7) of  Table \ref{tab_dxx_all} we present the total systematic error, $\sigma_{\rm sys}$, 
which is calculated from the quadratic sum of 
$\sigma_{\rm e}$, $\sigma_{\rm c}$ and 122 \ms\ we found from 21-cm analysis.
%
%
\begin{table*}
\caption{\dxx\ measured (in units of 10$^{-6}$) from different absorption systems}
\begin{center}
\begin{tabular}{lcccccccccccc}
\hline
\hline

Quasar & \zabs  & \multicolumn{7}{c}{UVES} & Keck & \multicolumn{2}{c}{dipole} & $\Delta$(\daa)\\
       &        & \dxx      & $\rm \sigma_{stat}$ & $\sigma_{\rm c}$ & $\sigma_{\rm e}$ & $\rm \sigma_{sys}$ & $\rm \sigma_{tot}$    & \daa     & \dxx & $\Theta$ & \daa & \\ [1.0ex] %
(1) & (2) & (3) & (4) &(5)&(6)&(7)&(8)&(9)&(10)&(11)&(12)&(13)\\
\hline
 J1623$+$0718  &  1.3356  &$-$3.7   &  3.0   &  1.0 & 1.2  &  1.6  &  3.4  &  $-$1.8$\pm$1.7 &....           &66.0  &$+$3.9$\pm$1.6 & $+$5.7$\pm$2.3 \\ 
\\                                                                  
 J2340$-$0053  &  1.3608  &$-$1.3   &  0.9   &  0.7 & 1.6  &  1.7  & 2.0   & $-$0.6$\pm$1.0  &$-$1.0$\pm$1.6 &90.6  &$-$0.1$\pm$1.2 & $+$0.5$\pm$1.6 \\ 
\\                                                                  
 J2358$-$1020  &  1.1730  &$+$1.8   &  1.8   &  0.8 & 1.8  &  2.0  & 2.7   & $+$0.9$\pm$1.4  &....           &84.9  &$+$0.8$\pm$1.1 & $-$0.1$\pm$1.8 \\ 
\\                                                                  
 J0501$-$0159  &  1.5605  &$+$3.0   &  2.1   &  0.6 & 2.2  &  2.3  & 3.1   & $+$1.5$\pm$1.6  &$+$4.7$\pm$10.9 &119.0 &$-$5.2$\pm$1.8 & $-$6.7$\pm$2.4 \\ 
\hline
  \multicolumn{1}{l}{}& \multicolumn{6}{c}{UVES} & \multicolumn{5}{c}{Keck}&\\
      \multicolumn{1}{l}{simple average}      & \multicolumn{6}{c}{$ $0.0 $\pm$1.5} & \multicolumn{5}{c}{$+$1.8 $\pm$2.8}\\
      \multicolumn{1}{l}{weighted average}    & \multicolumn{6}{c}{$-$0.1 $\pm$1.3} & \multicolumn{5}{c}{$-$0.9 $\pm$1.6}\\
\hline
\end{tabular}
\end{center}
\begin{flushleft}
Column 1: Object name. Column 2: absorption redshift. Columns 3-8: \dxx\ measurement, associated
statistical and systematic  errors (as discussed in Section 5) respectively. Column 9: \daa\ calculated 
based on the final values for \dxx\ assuming constancy of other constants. Column 10: 
\dxx\ measurements based on Keck/HIRES data. Column 11 and 12: angular distance in degrees between the 
quasar sight line and the best fitted dipole position from \citet{Webb11} and the predicted 
value for \daa\ based on dipole. Column 13: difference between our measurement and the prediction 
from dipole and its associated error. 
\end{flushleft}
\label{tab_dxx_all}
\end{table*}
%
%

\section{Results and Conclusions}
In Table~\ref{tab_dxx_all} we summarize the \dxx\ measurements in
individual systems. 
We recollect that the values are obtained under the assumption that 
the strongest UV and 21-cm absorption are produced by the same gas. 
The final errors in \dxx\ for our VLT/UVES measurements given in column (8) 
are the  quadratic sum of the statistical and systematic errors. 
We find the simple mean  of \dxx\ regardless of the associated errors in individual measurements to be 
$ $(0.0$\pm$1.5)$\times 10^{-6}$ with an rms of 3.0$\times 10^{-6}$ around the mean. 
A constant \dxx\ of $ $0.0$\times 10^{-6}$ has a reduced $\chi^2$ of 1.0  for our 
four UVES measurements that shows the estimated errors in \dxx\ are not underestimated. 
If we apply the standard procedure of weighting the data points by their inverse 
square errors (given in column (8) of Table~\ref{tab_dxx_all}), we get
$-(0.1\pm1.3)\times 10^{-6}$. 

\begin{figure} 
\centering
\includegraphics[width=0.75\hsize,bb=18 18 594 774,clip=,angle=90]{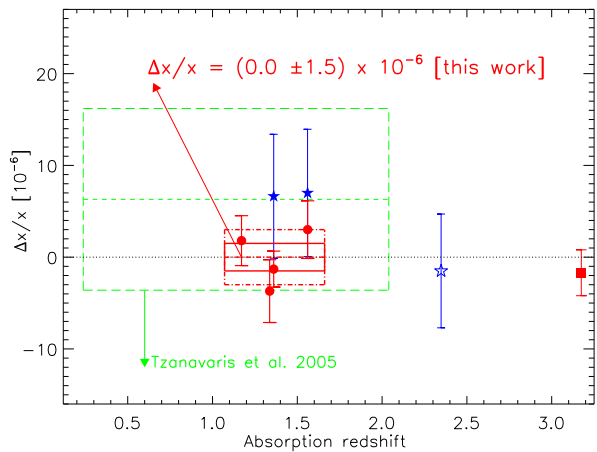}
\caption{Comparison of our \dxx\ measurements with three other measurements in the literature. 
Filled circles are our measurements and the filled square is from \citet{Srianand10} using 
a 21-cm absorber at $z \sim$ 3.174 along the line of sight of J1337$+$3152. 
The dashed-dotted box corresponds to the standard deviation of our four measurements
The 1$\sigma$ error around the mean is shown as a solid box. 
The  dashed line and  long-dashed box are the \dxx\ and its error measured by \citet{Tzanavaris07}.
The filled and empty stars are respectively from \citet{Kanekar10} and \citet{Kanekar06}. 
}
\label{fig_dx_all}
\end{figure}
%
Our VLBA images suggest that two of the quasars in the sample (i.e J1623$+$0718 and
J0501$-$0159) may have resolved structures at milliarcsec scale containing  more 
than 50\% of the flux. This may imply that  if the absorbing gas is extended then 
some additional 21-cm absorption can originate from the gas that is not probed by
the optical sight lines. However in the remaining two quasars this is not the case
as most of the flux is recovered in the unresolved VLBA component. If we only use
these two cases we derive \dxx\ = $+(0.2\pm1.6)\times10^{-6}$. 
This is very much consistent with what we find using all the four systems. 
Therefore the analysis presented here does not find any statistically significant variation in $x$
and this null result may not be related to systematics due to radio structure. 

There are two quasars for which we have spectra from both VLT and Keck. 
As can be seen from Table~\ref{tab_dxx_all} in both cases the VLT and Keck 
measurements are consistent with each other with Keck measurements having 
larger statistical uncertainties.
The mean \dxx\ from  VLT/UVES data of $ $(0.0$\pm$1.5)$\times 10^{-6}$  
is consistent with \dxx\ = $+$(1.8$\pm$2.8)$\times 10^{-6}$ from Keck/HIRES data where 
we could not include the systematic error.
This is also the case for the weighted means. 

In Fig. \ref{fig_dx_all} we compare our \dxx\ results with other \dxx\ measurements 
from the literature. Filled circles are our measurements. The dashed-dotted  
rectangular box indicates the mean and standard deviation of our measurements.
The solid rectangular box gives the mean and final error on it when combining the 4 measurements.
The green box with the dashed line gives the weighted mean and 1$\sigma$ found by \citet{Tzanavaris07}. 
The two filled stars are from \citet{Kanekar10} and the empty star is from \citet{Kanekar06}. 
The data point at $z\sim 3.174$ is from \citet{Srianand10}.

The better accuracy reached in our study is mainly due to the following reasons: 
(1) Systems are chosen to have narrow 21-cm absorption components.
(2) Three of the quasars have high resolution (R$\sim$45000) and 
high SNR UV-optical spectra obtained specifically for constraining \dxx\ with attached 
ThAr calibration lamps for each spectrum. This minimizes the systematic error of wavelength 
calibration. (3) Very high ($\sim$1 \kms per channel) or 
high ($2-4$ \kms per channel) resolution 21-cm spectra are used.
(4) As we could get repeated observations for 21-cm absorptions 
we are able to identify the RFI related problems in the absorption profiles.
 (5) We also estimate \dxx\ by using only absorbers for which the background sources are  
unresolved even at milliarcsec scale in VLBA images.
\begin{figure} 
\centering
\includegraphics[width=0.75\hsize,bb=18 18 594 774,clip=,angle=90]{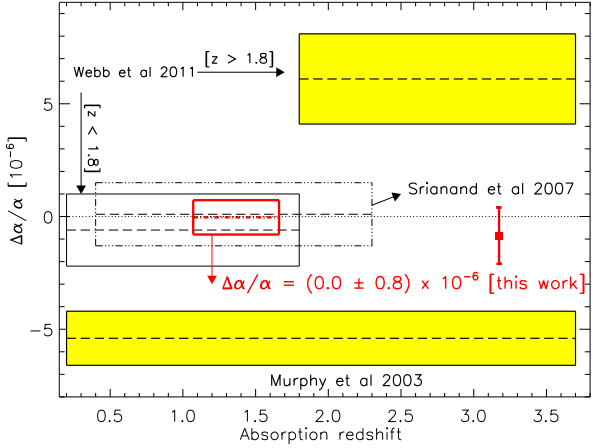}
\caption{Comparison of  \daa\ estimated in this work with other measurements in literature. 
The dashed dotted line and the surrounded solid box show our measured \daa\ and its error 
and the filled square is the one from 
\citet{Srianand10} based on the same method and assuming $\mu$ and $g_{\rm p}$ are not changing with time.}
\label{fig_alpha_all}
\end{figure}
\begin{figure} 
\centering
\includegraphics[width=0.75\hsize,bb=18 18 594 774,clip=,angle=90]{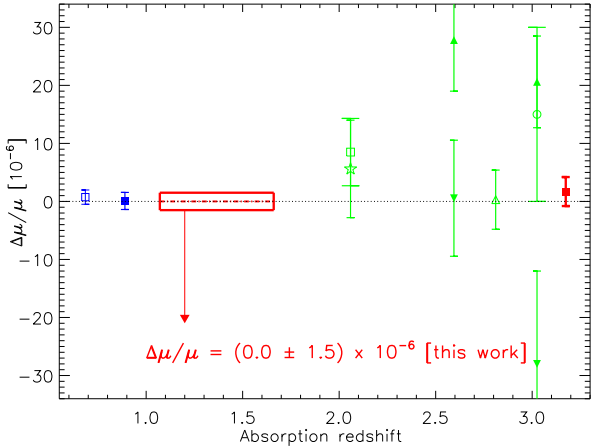}
\caption{Comparison of  \dmm\ estimated in this work with other measurements in literature. 
The dashed dotted line and the surrounded solid box show our measured \dmm\ and its error based 
on our \dxx\  measurements and assuming $\alpha$ and $g_p$ do not vary. Apart 
from the filled 
square at $z$$\sim$3.2 which is calculated from \dxx\ of \citet{Srianand10} with the 
assumption of non-variation of other constants, the rest of the measurements for  
$z > 2$ are based on the  analysis of $\rm H_2$  electronic transitions. 
The empty  triangle towards up is from \citet{King11}, 
empty star from  \citet{Malec10},
empty square with a large bar from \citet{vanWeerdenburg11},
empty  circle with a large bar from \citet{Wendt11},
filled triangles towards down from \citet{Thompson09b},
and filled triangles towards up from \citet{Reinhold06}.
The two   measurements at $z \le 1$ are based on the molecular inversion and rotational  
transitions. The filled and empty squares are from  \citet{Henkel08} and \citet{Murphy08} respectively.}
\label{fig_dmu_all}
\end{figure}

Although tight constraints on the variation of fundamental constants 
are obtained by comparing 21-cm and UV-optical redshifts,
the method is not exempt of systematics as all other methods in this field. 
As it is has already been discussed, the main source of uncertainty 
on \dxx\ is related to the assumption used to associate one of the several 
UV-optical components to the 21-cm absorption. By choosing absorbers with compact 
background radio sources at mas scale one can minimize the uncertainties related 
to the possibility that optical and radio sightlines are different. Different 
methods have been implemented to associate the UV-optical component to 
the 21-cm one. \citet{Tzanavaris07} associated the pixel with strongest UV-optical 
optical depth to the pixel with strongest 21-cm optical depth. 
In this work we follow the same idea but using 
components of simultaneous Voigt profile fitting models. 
Using the same method as \citet{Tzanavaris07},
we find \dxx\ = $(3.6\pm3.1)\times10^{-6}$. 
Keeping this in mind we will now discuss the implication of our constraint on \dxx\ on the 
variation of individual constants that constitute $x$.

As $x = g_{\rm p} \alpha^2/\mu$, its variation can be related to the variation of 
$g_{\rm p}$, $\alpha$, and/or $\mu$ via
$\Delta x/x = \Delta g_{\rm p}/g_{\rm p} + 2\times 
\Delta \alpha/\alpha-\Delta \mu/\mu$.
Therefore the constancy of $x$ 
can be related either to the constancy of all the three constants, $g_{\rm p}$, 
$\alpha$ and $\mu$, or to some complicated combination of variations of these 
constants with an overall null effect on $x$.
 
Assuming $\mu$ and $g_{\rm p}$ are constants then our measured \dxx\ translates 
to \daa\ = $\left(0.0\pm0.8\right)\times 10^{-6}$ which is one of the 
most stringent constraint on the variation of $\alpha$. In Fig~\ref{fig_alpha_all}
we summarize the available constraints on \daa\ from the literature.  It is 
clear that our measurements are consistent with \daa = $(0.1\pm1.5)\times10^{-6}$
(and a factor 2 better than that)
found by \citet{Srianand07b} and with the results of \citet{Webb11} for $z\le 1.8$
UVES data. Even if we use the conservative approach of \citet{Tzanavaris07} 
we get \daa\ = $(1.8\pm1.5)\times10^{-6}$ which is as good as the results from the 
MM method.

\citet{Webb11} used a combined set of absorbers observed with VLT/UVES and Keck/HIRES 
to conjecture about the possible presence of a spatial dipole pattern 
in the variation of $\alpha$. Their best fitted model indicates a spatial dipole in the 
direction with right ascension 17.5$\pm$0.9 h and declination $-58 \pm 9$ deg, significant 
at the 4.2$\sigma$ level with \daa\ = $A r \cos(\Theta)$, where 
$A = (1.1\pm0.25)\times10^{-6}\rm GLyr^{-1}$, $r(z)=ct(z)$, 
$c$ being the speed of light and $t(z)$ the look back time, and $\Theta$ is the angle 
on the sky between the quasar sight line and the best fit dipole position. 
We calculated the prediction of this model for our sample using        
the same cosmology as \citet{Webb11}. We also calculated the error on  \daa\ using
\begin{equation}
\sigma(\frac{\Delta\alpha}{\alpha}) = \sqrt{\left(\frac{\Delta\alpha/\alpha}{A}\right)^2\sigma_{A}^2+\left(\frac{\Delta\alpha/\alpha}{\cos(\Theta)}\right)^2\sigma_{\cos(\Theta)}^2}.
\end{equation}
The measured $\Theta$ and accordingly the predicted values for \daa\ 
are presented in columns 11 and 12  of Table~\ref{tab_dxx_all}. 
Last column of  Table~\ref{tab_dxx_all}  shows the  difference of \daa\ between 
the dipole and our  measurement  and its error. 
For the sight lines towards J1623$+$0718 and J0501$-$0159 the dipole predicts a variation 
of $\alpha$ at, respectively, the $\sim$2.4$\sigma$ and 2.9$\sigma$ level. In both cases,  our null results 
on the $\alpha$-variation are inconsistent with this dipole prediction at more than  90\% confidence level.
For the other  
two sight lines towards J2340$-$0053 and J2358$-$1020 our measurements are 
consistent with null variation as also predicted by the dipole. 
In the case of J1337$+$3152
\citet{Srianand10} measured \dxx\ = $-(1.7\pm1.7)\times10^{-6}$ for the absorber at \zabs\ = 3.174 
which translates to \daa\ = $-(0.9\pm0.9)\times10^{-6}$. The dipole prediction in this case is 
\daa\ = $-(2.70\pm1.9)\times10^{-6}$ with a difference of $-(1.80\pm2.1)\times10^{-6}$ 
with the measurement.   
The difference between these five measurements and the dipole predictions results in a 
$\chi^2$ of 14.5. The probability of $\chi^2$ $>$ 14.5  is $\sim$ 1\% 
which implies that the existence of a dipole is not favored by our measurements. 
More independent measurements especially towards systems where the dipole predicts 
large variations will be useful to confirm/refute the existence of the $\alpha-$dipole 
at higher significant level.

Assuming that $\alpha$ and $g_{\rm p}$ have been constant 
we derive \dmm\ = $\left(0.0 \pm1.5 \right)\times 10^{-6}$. 
Fig. \ref{fig_dmu_all} compares our results with other direct measurements of \dmm\  
obtained either using rotational transitions 
of $\rm H_2$ and HD molecules (for $z \ge 2.0$) or based on the comparison of 
NH$_3$ inversion transitions with some rotational transition lines 
(e.g. CO, CS, $\rm HC_3 N$; for $z \le 1.0$). 
While the constraints we get are not as good as the one obtained using NH$_3$  
they are very stringent compared to those based on H$_2$ at $z\ge2$. What is more interesting is
that our measurements fill the redshift gap between NH$_3$ and H$_2$ based
measurements (see Fig.~\ref{fig_dmu_all}).

If we use the 1$\sigma$ constraints on \daa\ found for $z\le1.8$ absorbers \citep[from][]{Srianand07b, Webb11}
and \dmm\ estimated at $z\sim0.7$ using NH$_3$ \citep{Kanekar11}, considering they are valid at $z\sim1.3$,  
we get $\Delta g_{\rm p}/g_{\rm p} \le 3.5 \times10^{-6} (1\sigma)$ from our \dxx\ measurements.

In summary, using 21-cm and metal UV absorption 
lines we are able to derive stringent constraints on the variation of $\alpha$, $\mu$, and $g_{\rm p}$. 
As discussed before the best estimate on \dmm\ at z $\le$1 is obtained 
by comparing the frequencies of NH$_3$  inversion transitions with rotational transitions 
of other molecules. The existing two measurements are towards the line of sight of 
two well known gravitationally lensed BL Lacs (B~0218+357 and PKS~1830$-$211) that 
show complex radio morphologies. As different transitions 
occur at different frequencies the dependence of the background radio structure on frequency is
an important source of systematic error \citep[see][]{Murphy08,Kanekar11}. Therefore 
detecting NH$_3$ and other molecules towards unlensed compact radio sources is important to constrain
\dmm. Unlike NH$_3$ and other complex heavy molecules, 21-cm absorption is 
more frequently detected towards normal radio sources covering a wide redshift range. The main
source of systematics in this method is related to how accurately the 21-cm
absorption component is associated to the corresponding metal line component. 
More measurements towards compact radio sources are needed to address this issue 
adequately. Future blind searches for 21-cm absorption 
using the upcoming Square Kilometer Array (SKA) path finders hopefully will provide 
a large number of suitable targets to perform such measurements.
\section*{acknowledgement}
We would like to thank the anonymous referee for the helpful and constructive comments.  
We thank GMRT, GBT, and VLBA staff for their support during the observation. GMRT is an 
international facility run by the National Centre for Radio Astrophysics 
of the Tata Institute of Fundamental Research. GBT and VLBA are run by National Radio Astronomy Observatory. 
The VLBA data from 2010 were correlated using NRAO's implementation of the DiFX software 
correlator that was developed as part of the Australian Major National Research facilities 
Programme and operated under license. The National Radio Astronomy Observatory is a facility of 
the National Science Foundation operated under cooperative agreement by Associative 
Universities, Inc.  
R. S. and P. P. J. gratefully acknowledge support 
from the Indo-French Centre for the Promotion of Advanced Research (Centre Franco-Indian pour la Promotion de la
Recherche Avanc\'ee) under contract No. 4304-2.
%
%
%
\def\aj{AJ}%
\def\actaa{Acta Astron.}%
\def\araa{ARA\&A}%
\def\apj{ApJ}%
\def\apjl{ApJ}%
\def\apjs{ApJS}%
\def\ao{Appl.~Opt.}%
\def\apss{Ap\&SS}%
\def\aap{A\&A}%
\def\aapr{A\&A~Rev.}%
\def\aaps{A\&AS}%
\def\azh{AZh}%
\def\baas{BAAS}%
\def\bac{Bull. astr. Inst. Czechosl.}%
\def\caa{Chinese Astron. Astrophys.}%
\def\cjaa{Chinese J. Astron. Astrophys.}%
\def\icarus{Icarus}%
\def\jcap{J. Cosmology Astropart. Phys.}%
\def\jrasc{JRASC}%
\def\mnras{MNRAS}%
\def\memras{MmRAS}%
\def\na{New A}%
\def\nar{New A Rev.}%
\def\pasa{PASA}%
\def\pra{Phys.~Rev.~A}%
\def\prb{Phys.~Rev.~B}%
\def\prc{Phys.~Rev.~C}%
\def\prd{Phys.~Rev.~D}%
\def\pre{Phys.~Rev.~E}%
\def\prl{Phys.~Rev.~Lett.}%
\def\pasp{PASP}%
\def\pasj{PASJ}%
\def\qjras{QJRAS}%
\def\rmxaa{Rev. Mexicana Astron. Astrofis.}%
\def\skytel{S\&T}%
\def\solphys{Sol.~Phys.}%
\def\sovast{Soviet~Ast.}%
\def\ssr{Space~Sci.~Rev.}%
\def\zap{ZAp}%
\def\nat{Nature}%
\def\iaucirc{IAU~Circ.}%
\def\aplett{Astrophys.~Lett.}%
\def\apspr{Astrophys.~Space~Phys.~Res.}%
\def\bain{Bull.~Astron.~Inst.~Netherlands}%
\def\fcp{Fund.~Cosmic~Phys.}%
\def\gca{Geochim.~Cosmochim.~Acta}%
\def\grl{Geophys.~Res.~Lett.}%
\def\jcp{J.~Chem.~Phys.}%
\def\jgr{J.~Geophys.~Res.}%
\def\jqsrt{J.~Quant.~Spec.~Radiat.~Transf.}%
\def\memsai{Mem.~Soc.~Astron.~Italiana}%
\def\nphysa{Nucl.~Phys.~A}%
\def\physrep{Phys.~Rep.}%
\def\physscr{Phys.~Scr}%
\def\planss{Planet.~Space~Sci.}%
\def\procspie{Proc.~SPIE}%
\let\astap=\aap
\let\apjlett=\apjl
\let\apjsupp=\apjs
\let\applopt=\ao
\bibliographystyle{mn}
\bibliography{mybib.bib}
%
\appendix
\section{Results of correlation analysis}
\begin{table*}
\caption{Shifts of individual spectra relative to the combined one 
in J0108$-$0037 at the position of different absorption profiles}
\begin{center}
\begin{tabular}{lcccc}
\hline
\hline
species & EXP1  & EXP2  & EXP3  & EXP4 \\
        & (\ms) & (\ms) & (\ms) & (\ms) \\
~~~~~~~~(1) & (2)    & (3) &  (4)  \\
\hline
\NiII\ $\lambda$1454 &   656$\pm$ 505 &  -826$\pm$ 681 &   531$\pm$ 458 &  -536$\pm$ 463 \\
\NiII\ $\lambda$1467 &  -131$\pm$ 595 & -1081$\pm$ 835 &  -973$\pm$ 764 &   484$\pm$ 534 \\
\NiII\ $\lambda$1502 & -1094$\pm$ 556 &   925$\pm$ 887 &    62$\pm$ 816 &    65$\pm$ 941 \\
\NiII\ $\lambda$1703 &  -391$\pm$ 359 &   169$\pm$ 473 &  1306$\pm$ 778 &   161$\pm$ 618 \\
\NiII\ $\lambda$1709 &   -11$\pm$ 184 &   230$\pm$ 211 &  -408$\pm$ 227 &    43$\pm$ 195 \\
\NiII\ $\lambda$1741 &  -145$\pm$ 142 &   275$\pm$ 164 &  -160$\pm$ 171 &   -98$\pm$ 162 \\
\NiII\ $\lambda$1751 &  -496$\pm$ 204 &   240$\pm$ 221 &   165$\pm$ 223 &  -152$\pm$ 211 \\
\CrII\ $\lambda$2056 &  -149$\pm$ 109 &   109$\pm$ 121 &   133$\pm$ 135 &   -80$\pm$ 121 \\
\CrII\ $\lambda$2062 &   -83$\pm$ 117 &  -168$\pm$ 130 &    99$\pm$ 139 &   151$\pm$ 132 \\
\CrII\ $\lambda$2066 &   -66$\pm$ 162 &    15$\pm$ 176 &  -267$\pm$ 202 &   220$\pm$ 179 \\
\MnII\ $\lambda$2576 &    32$\pm$ 102 &   115$\pm$ 111 &   -44$\pm$ 110 &  -123$\pm$ 105 \\
\MnII\ $\lambda$2594 &  -176$\pm$ 133 &  -220$\pm$ 151 &   181$\pm$ 150 &   156$\pm$ 142 \\
\MnII\ $\lambda$2606 &  -187$\pm$ 149 &   607$\pm$ 173 &  -181$\pm$ 168 &  -140$\pm$ 165 \\
\ZnII\ $\lambda$2026 &  -207$\pm$ 166 &   234$\pm$ 188 &  -209$\pm$ 194 &   170$\pm$ 184 \\
\ZnII\ $\lambda$2062 &  -108$\pm$ 166 &    79$\pm$ 175 &  -267$\pm$ 198 &   218$\pm$ 182 \\
\FeII\ $\lambda$1611 &  -282$\pm$ 228 &    74$\pm$ 290 &   752$\pm$ 305 &  -373$\pm$ 267 \\
\FeII\ $\lambda$2249 &  -145$\pm$ 103 &    80$\pm$ 110 &  -118$\pm$ 116 &   182$\pm$ 118 \\
\FeII\ $\lambda$2260 &  -358$\pm$  90 &   300$\pm$ 100 &   -56$\pm$ 105 &   183$\pm$ 106 \\
\SiII\ $\lambda$1808 &  -237$\pm$ 108 &   132$\pm$ 127 &   239$\pm$ 138 &   -50$\pm$ 124 \\
\hline
weighted mean               &  -171 & +127  &  -14  & +38     \\
weighted standard deviation &   152 &  207  &   223 & 166     \\
\multicolumn{1}{l}{weighted standard deviation of all exposures}       &   \multicolumn{4}{c}{134}   \\  
\hline
\end{tabular}
\end{center}
\begin{flushleft}
\end{flushleft}
\label{tab0108shift_opt}
\end{table*}

\begin{table*}
\caption{Shifts of individual spectra  relative to the combined one 
in J1623$+$0718 at the position of different absorption profiles}
\begin{center}
\begin{tabular}{lcccc}
\hline
\hline
species & EXP1  & EXP2  & EXP3  & EXP4     \\
        & (\ms) & (\ms) & (\ms) & (\ms)  \\
\hline
\NiII\ $\lambda$1454 &  2075$\pm$ 573 &  -414$\pm$ 582 &    94$\pm$ 393 &  -840$\pm$ 364 \\
\NiII\ $\lambda$1703 &  2230$\pm$1013 &  -531$\pm$ 682 &   378$\pm$ 446 &  -414$\pm$ 419 \\
\NiII\ $\lambda$1709 &  -161$\pm$ 465 &   520$\pm$ 414 &   485$\pm$ 499 & -1040$\pm$ 414 \\
\CrII\ $\lambda$2062 &   177$\pm$ 564 &  -180$\pm$ 341 &  -191$\pm$ 530 &   242$\pm$ 274 \\
\CrII\ $\lambda$2066 &  -434$\pm$ 687 &   461$\pm$ 633 &   335$\pm$ 745 &  -684$\pm$ 602 \\
\ZnII\ $\lambda$2062 & -1434$\pm$ 643 &   461$\pm$ 633 &    72$\pm$ 615 &   460$\pm$ 408 \\
\MnII\ $\lambda$2576 &   498$\pm$ 426 &     0$\pm$ 346 &   235$\pm$ 420 &  -217$\pm$ 224 \\
\MnII\ $\lambda$2594 &   -46$\pm$ 458 &    80$\pm$ 303 &   572$\pm$ 386 &  -185$\pm$ 213 \\
\MnII\ $\lambda$2606 &   475$\pm$ 481 & -1068$\pm$ 735 & -1073$\pm$ 574 &   194$\pm$ 283 \\
\FeII\ $\lambda$1611 &  2002$\pm$ 612 &  1394$\pm$ 770 &   766$\pm$1559 &  -432$\pm$ 335 \\
\FeII\ $\lambda$2249 &  -555$\pm$ 389 &   500$\pm$ 273 &  -415$\pm$ 424 &   -70$\pm$ 305 \\
\FeII\ $\lambda$2260 &   342$\pm$ 320 &   635$\pm$ 322 &  -632$\pm$ 366 &  -164$\pm$ 194 \\
\FeII\ $\lambda$2374 &   -55$\pm$ 184 &   678$\pm$ 161 &  -913$\pm$ 193 &     8$\pm$ 110 \\
\FeII\ $\lambda$2382 &  -275$\pm$ 142 &   460$\pm$ 139 &  -329$\pm$ 160 &    25$\pm$  89 \\
\FeII\ $\lambda$2586 &   196$\pm$ 160 &   451$\pm$ 139 &  -237$\pm$ 158 &  -307$\pm$  87 \\
\SiII\ $\lambda$1808 &   142$\pm$ 261 &   224$\pm$ 220 &   414$\pm$ 253 &  -381$\pm$ 180 \\
\hline
weighted mean               &  +44 & +401  & -233  &  -145    \\
weighted standard deviation &  543  &314    &  480  &  251    \\
\multicolumn{1}{l}{weighted standard deviation of all exposures}       &   \multicolumn{4}{c}{313}   \\  
\hline
\end{tabular}
\end{center}
\begin{flushleft}
\end{flushleft}
\label{tab1623shift_opt}
\end{table*}

\begin{table*}
\caption{Shifts of individual spectra relative to the combined one 
in J2340$-$0053 at the position of different absorption profiles}
\begin{center}
\begin{tabular}{lccccccccc}
\hline
\hline
species & EXP1  & EXP2  & EXP3  & EXP4  & EXP5  & EXP6  & HIRES \\
        & (\ms) & (\ms) & (\ms) & (\ms) & (\ms) & (\ms) & (\ms) \\
\hline
\NiII\ $\lambda$1709 &  -645$\pm$ 358 &   212$\pm$ 428 &   144$\pm$ 336 &   759$\pm$ 455 &   380$\pm$ 412 &    74$\pm$ 345 & 2530$\pm$ 296\\
\NiII\ $\lambda$1741 &   -81$\pm$ 380 &     5$\pm$ 361 &   447$\pm$ 296 &   668$\pm$ 321 &    97$\pm$ 282 &  -958$\pm$ 317 &  626$\pm$ 232\\
\NiII\ $\lambda$1751 &  -409$\pm$ 513 &  -122$\pm$ 392 &   588$\pm$ 297 &  -180$\pm$ 373 &  -406$\pm$ 354 &   142$\pm$ 332 &   65$\pm$ 521\\
\CrII\ $\lambda$2056 &  -169$\pm$ 759 &   303$\pm$ 868 &   139$\pm$ 761 &  -307$\pm$ 922 &   263$\pm$ 811 &   -41$\pm$ 555 &  -43$\pm$ 196\\
\CrII\ $\lambda$2062 &  -206$\pm$ 749 &   184$\pm$ 747 &   217$\pm$ 574 &    49$\pm$ 927 &  -213$\pm$ 574 &  -467$\pm$ 606 &  244$\pm$ 268\\
\CrII\ $\lambda$2066 &   728$\pm$1026 &  -646$\pm$1201 &   233$\pm$ 988 &  -561$\pm$1246 &    74$\pm$1179 &  -446$\pm$1042 &   ----       \\
\ZnII\ $\lambda$2026 &   155$\pm$ 425 &   130$\pm$ 486 &   -88$\pm$ 377 &  -692$\pm$ 389 &   205$\pm$ 290 &   194$\pm$ 325 &   ----       \\
\MnII\ $\lambda$2576 &  -197$\pm$ 351 &   219$\pm$ 407 &   487$\pm$ 400 &   -10$\pm$ 384 &   -13$\pm$ 368 &  -209$\pm$ 290 &   ----       \\
\MnII\ $\lambda$2594 &  -621$\pm$ 375 &   129$\pm$ 370 &   122$\pm$ 384 &   -41$\pm$ 363 &   476$\pm$ 361 &  -110$\pm$ 274 &   ----       \\
\MnII\ $\lambda$2606 &  -143$\pm$ 317 &  -439$\pm$ 405 &   885$\pm$ 310 &  -139$\pm$ 361 &   436$\pm$ 298 & -1199$\pm$ 284 &   ----       \\
\FeII\ $\lambda$1608 &  -309$\pm$  66 &   117$\pm$  72 &   275$\pm$  62 &   -48$\pm$  81 &   113$\pm$  68 &  -205$\pm$  65 &  47$\pm$  48 \\
\FeII\ $\lambda$1611 &   -15$\pm$ 408 &   128$\pm$ 385 &    66$\pm$ 287 &   281$\pm$ 493 &  -325$\pm$ 532 &  -764$\pm$ 456 & 2271$\pm$ 539\\
\FeII\ $\lambda$2249 &  -103$\pm$ 184 &  -104$\pm$ 260 &  -221$\pm$ 198 &  -307$\pm$ 195 &   171$\pm$ 183 &   138$\pm$ 149 & -433$\pm$ 193\\
\FeII\ $\lambda$2260 & -1272$\pm$ 138 &   208$\pm$ 133 &   110$\pm$ 117 &    43$\pm$ 136 &   594$\pm$ 109 &   132$\pm$ 101 &  -27$\pm$ 21 \\
\FeII\ $\lambda$2374 &  -308$\pm$  41 &   185$\pm$  41 &   230$\pm$  37 &    35$\pm$  40 &   315$\pm$  35 &  -294$\pm$  32 &  148$\pm$  57\\
\SiII\ $\lambda$1808 &  -380$\pm$ 139 &  -137$\pm$ 160 &   357$\pm$ 144 &   122$\pm$ 154 &   249$\pm$ 157 &  -165$\pm$ 134 & -229$\pm$  92\\
\hline
weighted mean               &  -352 &  146  &   236 &  18   & 283   &  -238 &  110  \\
weighted standard deviation &  296  &  127  &   148 &  173  & 184   &   238 &  438  \\
\multicolumn{1}{l}{weighted standard deviation of all exposures}       & \multicolumn{6}{c}{204$^\star$}&          \\  
\hline
\end{tabular}
\end{center}
\begin{flushleft}

$\star$standard deviation in HIRES column is not included in the averaged standard deviation.
\end{flushleft}
\label{tab2340shift_opt}
\end{table*}

\begin{table*}
\caption{Shifts of individual spectra  relative to the combined one 
in J2358$-$1020 at the position of different absorption profiles}
\begin{center}
\begin{tabular}{lccccccc}
\hline
\hline
species & EXP1  & EXP2  & EXP3  & EXP4  & EXP5  & EXP6  & EXP7    \\
        & (\ms) & (\ms) & (\ms) & (\ms) & (\ms)& (\ms)& (\ms)  \\
\hline
\ZnII\ $\lambda$2026 &  -509$\pm$ 549 &   -477$\pm$ 376 &    -744$\pm$ 629 &  1083$\pm$ 395 & 173$\pm$ 266 &  165$\pm$ 326 &  -511$\pm$ 553 \\
\MnII\ $\lambda$2576 &  -555$\pm$ 521 &     73$\pm$ 419 &    -658$\pm$ 420 &  -488$\pm$ 406 & 150$\pm$ 317 &  322$\pm$ 413 &  1316$\pm$ 578 \\
\FeII\ $\lambda$2249 &   -13$\pm$ 747 &   -317$\pm$1013 &    -370$\pm$ 596 &  -203$\pm$ 452 & 753$\pm$ 730 &  397$\pm$ 456 &  -568$\pm$ 473 \\
\FeII\ $\lambda$2344 &   -19$\pm$ 234 &     88$\pm$ 251 &    -257$\pm$ 247 &  -311$\pm$ 189 &  32$\pm$ 151 &  290$\pm$ 201 &    72$\pm$ 201 \\
\FeII\ $\lambda$2374 &  -800$\pm$ 294 &  - 118$\pm$ 213 &     -35$\pm$ 261 &    10$\pm$ 177 & 205$\pm$ 156 &  188$\pm$ 174 &   392$\pm$ 264 \\
\FeII\ $\lambda$2586 &  -130$\pm$ 268 &    331$\pm$ 243 &      24$\pm$ 245 &  -213$\pm$ 166 & 139$\pm$ 165 &  258$\pm$ 180 &  -178$\pm$ 246 \\
\SiII\ $\lambda$1808 &   -20$\pm$ 577 &   -619$\pm$ 661 &   -1897$\pm$ 719 &  -497$\pm$ 499 & 241$\pm$ 421 &  612$\pm$ 527 &  1975$\pm$ 768 \\
\hline
weighted mean               &  -275  & +70  &  -242 & -129  & +142  & +147   &  +111 \\
weighted standard deviation &   357  &  288  &  431  &  374  &113    & 260   &  561  \\
\multicolumn{1}{l}{weighted standard deviation of all exposures}       &    \multicolumn{6}{c}{170}  \\  
\hline
\end{tabular}
\end{center}
\begin{flushleft}
\end{flushleft}
\label{tab2358shift_opt}
\end{table*}
\end{document}